\documentclass[twocolumn]{aastex62}

\usepackage{natbib}

\newcommand{\Swift}{{\it Swift}}

\newcommand{\timp}{t_{\rm imp}}
\newcommand{\days}{{\rm days}}
\newcommand{\kms }{{\rm km~s^{-1}}}
\newcommand{\lumL}{\mathcal{L}}
\newcommand{\Lp}{\lumL_{\nu,{\rm p}}}
\newcommand{\Msun}{M_\odot}
\newcommand{\Lhal}{\mathcal{L}_{\rm H\alpha}}
\newcommand{\rhocsm}{\rho_{\rm csm}}

\newcommand{\densu}{{\rm g~cm^{-3}}}
\newcommand{\csm}{{\rm csm}}
\newcommand{\Hal}{H$\alpha$}
\newcommand{\CaII}{\ion{Ca}{2}}

\newcommand{\MgII}{\ion{Mg}{2}}

\newcommand{\affilCal}{Department of Astronomy, University of California, Berkeley, CA 94720-3411, USA}
\newcommand{\affilLBL}{Lawrence Berkeley National Laboratory, 1 Cyclotron Road, MS 50B-4206, Berkeley, CA 94720, USA}
\newcommand{\affilHUJ}{Racah Institute of Physics, The Hebrew University of Jerusalem, Jerusalem, 91904, Israel}
\newcommand{\affilOxford}{Astrophysics, Department of Physics, University of Oxford, Denys Wilkinson Building, Keble Road, Oxford OX1 3RH, UK}
\newcommand{\affilUW}{Department of Astronomy, University of Washington, Box 351580, U.W., Seattle, WA 98195-1580, USA}
\newcommand{\affilQueens}{Astrophysics Research Centre, School of Mathematics and Physics, Queens University Belfast, Belfast BT7 1NN, UK}
\newcommand{\affilSouthampton}{Department of Physics and Astronomy, University of Southampton, Southampton SO17 1BJ, UK}
\newcommand{\affilASU}{School of Earth and Space Exploration, Arizona State University, Tempe, AZ 85287, USA}
\newcommand{\affilDavis}{Department of Physics, University of California, Davis, 1 Shields Ave, Davis, CA 95616, USA}
\newcommand{\affilMiller}{Miller Senior Fellow, Miller Institute for Basic Research in Science, University of California, Berkeley, CA 94720, USA}
\newcommand{\affilSTScI}{Space Telescope Science Institute, 3700 San Martin Drive, Baltimore, MD 21218, USA}
\newcommand{\affilStock}{The Oskar Klein Centre, Department of Physics, Stockholm University, AlbaNova, SE-106 91 Stockholm, Sweden}
\newcommand{\affilUMN}{School of Physics \& Astronomy, University of Minnesota, 116 Church Street S.E., Minneapolis, MN 55455, USA}

\shorttitle{The Circumstellar Environment of SN~2015cp}
\shortauthors{Harris, C.~E., et al.}

\begin{document}


  \title{Don't Blink: Constraining the Circumstellar Environment of the Interacting Type~Ia Supernova 2015cp}

	\author[0000-0002-1751-7474]{C.~E.~Harris}
    \affiliation{\affilCal}
    \affiliation{\affilLBL}

	\author[0000-0002-3389-0586]{P.~E.~Nugent}
    \affiliation{\affilCal}
    \affiliation{\affilLBL}
  
	\author[0000-0002-5936-1156]{A.~Horesh}
	\affiliation{\affilHUJ}

	\author[0000-0002-7735-5796]{J.~S.~Bright}
    \affiliation{\affilOxford}

	\author{R.~P.~Fender}
    \affiliation{\affilOxford}

	\author{M.~L.~Graham}
    \affiliation{\affilCal}
    \affiliation{\affilUW}

	\author{K.~Maguire}
    \affiliation{\affilQueens}

	\author{M.~Smith}
    \affiliation{\affilSouthampton}

	\author{N.~Butler}
    \affiliation{\affilASU}

	\author[0000-0001-8818-0795]{S.~Valenti}
    \affiliation{\affilDavis}

	\author{A.~V.~Filippenko}
    \affiliation{\affilCal}
    \affiliation{\affilMiller}

	\author{O.~Fox}
    \affiliation{\affilSTScI}

	\author[0000-0002-4163-4996]{A.~Goobar}
    \affiliation{\affilStock}

	\author{P.~L.~Kelly}
    \affiliation{\affilCal}
    \affiliation{\affilUMN}

	\author{K.~J.~Shen}
    \affiliation{\affilCal}
    
    \correspondingauthor{Chelsea E. Harris}
    \email{chelseaharris@lbl.gov}


\begin{abstract}    
	Despite their cosmological utility, the progenitors of 
    Type~Ia supernovae (SNe~Ia) are still unknown, with many 
    efforts focused on whether accretion from a nondegenerate
    companion can grow a carbon-oxygen white dwarf to near the 
    Chandrasekhar mass.
    The association of SNe~Ia resembling SN~1991T (``91T-like'') 
    with circumstellar interaction may be evidence for this 
    ``single-degenerate'' channel.
    However, the observed circumstellar medium (CSM) in these 
    interacting systems is unlike a stellar wind -- of particular 
    interest, it is sometimes detached from the stellar surface, 
    residing at $\sim 10^{16}~{\rm cm}$.
	A {\it Hubble Space Telescope (HST)} program to discover detached CSM 
    around 91T-like SNe~Ia successfully discovered interaction nearly 
    two years after explosion in SN~2015cp \citep{Graham+18}. 
    In this work, we present radio and X-ray follow-up observations of 
    SN~2015cp and analyze them in the framework of \citet{HNK16} to 
    limit the properties of a constant-density CSM shell in this system.
	Assuming the {\it HST} detection was shortly after the shock crossed the 
    CSM, we constrain the total CSM mass in this system to be $< 0.5~\Msun$.
    This limit is comparable to the CSM mass of supernova PTF11kx, 
    but does not rule out lower masses predicted for recurrent novae.
    From lessons learned modeling PTF11kx and SN~2015cp, we suggest a 
    strategy for future observations of these events to increase the 
    sample of known interacting SNe~Ia.
\end{abstract}


\section{Introduction}
 
Broadly speaking, Type~Ia supernovae (SNe~Ia) are hydrogen-deficient 
thermonuclear explosions of white dwarfs 
\citep[see, e.g.,][for a review of SNe and their optical spectra]{Filippenko97}. 
The landscape of the debate regarding the detailed nature of Type~Ia supernova (SN~Ia) progenitors has not changed much 
since the 1980s, despite the use of SNe~Ia as increasingly precise cosmological tools. 
The review by \citet{Branch+95}, which concludes that
``the coalescence of pairs of [carbon-oxygen] white dwarfs, 
and the accretion of hydrogen on a thermal time scale via Roche-lobe overflow 
from subgiant donors, are the two most promising candidate progenitor mechanisms 
for SNe~Ia,'' and ``there is no strong objection to the notion that several,
or even all, candidates contribute,''
still largely holds today. The first scenario involving
two carbon-oxygen white dwarfs (CO WDs) is typically referred to as the 
``double-degenerate'' (DD) channel, and the second which involves a
nondegenerate companion is the ``single-degenerate'' (SD) channel. 
Two significant changes since this review are that the helium-shell detonation
(``double-detonation'') variant of the DD channel has been brought back into 
the mainstream \citep[e.g.,][]{Fink+10,ShenMoore14}, 
and some authors suggest a variant of the double-degenerate scenario in 
which the CO WD merges with the degenerate core of an asymptotic giant 
branch (AGB) companion \citep[e.g.,][]{Soker13}. 

Generally, though, efforts have been focused on distinguishing whether 
SD or DD is the dominant channel for forming SNe~Ia.
One of the identifying characteristics of the SD channel is that it
can create a dense, extended, hydrogen-rich
circumstellar medium (CSM), while DD companions do not ---
hydrogen-rich material from DD channels is swept over within a few days, or 
is quite distant 
($\gtrsim 10^{17}~{\rm cm}$) and low density \citep[e.g.][]{RaskinKasen13,Shen+13}.
From deep radio limits searching for SN ejecta interaction with a red-giant (RG) wind, 
\citet{Chomiuk+16} constrain the occurrence of RG companions for normal 
SNe~Ia to $<10\%$.
Yet there is evidence that {\it some} SNe~Ia have circumstellar gas:
a small number of SNe~Ia,  called ``Type~Ia-CSM'' by \citet{Silverman+13} who
characterized the population, are observed to interact 
with extremely dense CSM. 
As noted by \citet{Silverman+13} and \citet{Leloudas+15}, all are spectroscopically 
like SN~1991T \citep{Filippenko+92} or SN~1999aa \citep{Garavini+04},
i.e., they have strong 
\ion{Fe}{3} and weak \ion{Si}{2} lines near maximum light \citep{Branch+93}.
Hereafter, we will use the term ``91T-like'' to mean resembling either of these
SNe, rather than requiring the very weak \ion{Ca}{2}~H\&K absorption characteristic 
of SN~1991T itself.
The association of interaction with only this subgroup, which is also
associated with younger stellar populations, raises the question of 
whether 91T-like SNe~Ia have different progenitors compared to
normal SNe~Ia --- they may represent a SD channel for forming SNe~Ia.

Though not distinguished by \citet{Silverman+13}, 
interacting SNe actually fall into two categories:
those with prompt interaction, and those with delayed interaction.
In the SN~Ia-CSM class, the two instances of the latter scenario are
SN 2002ic \citep{WoodVasey04} and PTF11kx \citep{Dilday+12}.
The transition of PTF11kx from normal to interacting was well observed 
both spectroscopically and photometrically. 
Such transitions have also been reported in a few 
SNe Ib \citep{Milisav+15,Mauerhan+18},
and SN 1987A provides a famous, though extreme, example 
from the Type~II class \citep{Larsson+11,Fransson+15}.
To encompass the delayed interaction group, we use the label 
Type~X;n SNe --- for example,
SN 2014C is an SN Ib;n \citep{Milisav+15}, and PTF11kx is an SN~Ia;n.
The ``;n'' label captures the observational
properties of the class 
(the SN-only and interacting phases are independent, observationally) 
and allows for the application to different SN types.
Note that the ``n'' label indicates narrow emission lines, 
in accordance with the canonical interaction class (Type IIn), 
and that in this case ``narrow'' is relative to SN lines --- i.e., 
the narrow lines are $\lesssim 5,000~\kms$ and not {\it necessarily} unresolved nor tracing pre-shock CSM.

There are three important reasons to distinguish SNe X;n from promptly
interacting SNe that surmount the abhorrence of increasing SN taxonomic entropy.
First, that clearly the CSM in an SN X;n is 
likely moulded by different physical processes 
than that of prompt interactors (PTF11kx is an outlier in the SN~Ia-CSM class).
Moreover, it is crucial leverage that
the underlying SN type in SNe X;n can be unambiguously
classified and its explosion energetics inferred
from pre-interaction spectra and light curves.
Finally, distinguishing these events is necessary because studies of 
SNe X;n require new and unique methods for their 
discovery, classification, and analysis --- for example, the 
asymptotic hydrodynamic solutions from \citet{Chevalier82a} do not apply 
to these systems.

Possibly representing SNe~Ia;n with the
longest delay between explosion and interaction
are those SNe~Ia with time-variable narrow absorption lines.
Such lines in the near-maximum-light spectra heralded interaction for PTF11kx, 
and individual SNe~Ia show evidence for time-variable lines, though much
weaker than in the case of PTF11kx \citep{Patat+07,Simon+09}.
In statistical analyses it has been shown that 
$\sim 20$\% of SNe~Ia in S0 or later-type galaxies
have time-variable or blueshifted narrow Na~I~D absorption 
features \citep{Sternberg+11,Maguire+13,Sternberg+14}.
Whether these time-variable and blueshifted absorption features,
unaccompanied by later interaction, arise from CSM at $\sim 10^{17}~{\rm cm}$
or unassociated and perhaps more distant interstellar gas is still debated 
\citep{Chugai08,Wang+08,Borkowski+09,Bulla+18},
stoked by the fact that these events prefer dusty and gas-rich 
host galaxies (though the SNe themselves are not necessarily 
heavily extinguished).
If CSM is the origin of these features, interaction would not happen for years 
and typical SN~Ia observations --- which only capture the light curve for $\sim$100 days
near maximum brightness --- would not see the interaction.

To measure how frequently 91T-like SNe~Ia interact at late times, 
the {\it Hubble Space Telescope (HST)} Snapshot survey GO-14779 (PI\ M. L.  Graham) surveyed 71 SNe with ages of 1--3 yr (nearly all of them SNe~Ia, with some SNe~IIn) 
throughout the year 2017 in the near-ultraviolet (NUV).
The target list of 80 objects prioritized 91T-like SNe and those 
with blueshifted Na~I~D absorption.
This program discovered delayed interaction in 
SN 2015cp (also known as PS15dpq and PTF15fel), confirmed by the presence of broad 
\Hal\ emission in a follow-up optical spectrum. 
From the photometric fit, SN 2015cp exploded on 2015 November 1
(approximate day of first light) and thus the NUV detection was 
on day 681 after explosion.
In this paper, all times are reported relative
to the day of explosion unless stated otherwise.
The only pre-interaction optical spectrum available for this event was its 
classification spectrum 60 days after explosion, which shows no signs of CSM.
Details of the full survey and the NUV/optical observations 
of SN 2015cp are presented by \citet{Graham+18}.

In this work we report radio and X-ray follow-up observations of 
SN 2015cp and analyze the radio data, using the constant-density
shell models of \citet[][hereafter ``HNK16'']{HNK16} 
to constrain the total mass of CSM and other properties.
An overview of the HNK16 models is provided in \S\ref{sec:mod_desc}.
In \S\ref{sec:obs} we present the observations of SN 2015cp taken 
$\sim 80$ days after the NUV detection with the Arcminute Microkelvin Imager (AMI), 
the Jansky Very Large Array (VLA), 
and the Neil Gehrels {\it Swift} Observatory.
We analyze the radio nondetections in
\S\ref{sec:analysis} to place upper limits on the CSM mass, extent, and density.
In \S\ref{sec:conc} we summarize our results, 
contextualize the constraints with SN~Ia progenitor theory,
and outline how to systematically find and characterize SNe~Ia;n.

\section{Interaction Model Summary}
\label{sec:mod_desc}

HNK16 addressed the scenario of SN Ia ejecta impacting 
a constant-density, distant, finite-extent shell of CSM. 
Though we encourage a familiarity with the synchrotron radio light-curve 
behavior described by HNK16,
we will summarize the main conclusions of that work relevant to this study.
For more details on these equations, including derivations and normalizations, 
we refer to HNK16.
A glossary of variables is provided in Table~\ref{tab:gloss}.

\begin{deluxetable*}{llll}
  \tablecolumns{4}
  \tablecaption{Glossary of Variables
                \label{tab:gloss}}
  \tablehead{\colhead{Symbol} & \colhead{Description} & \colhead{Units in this work} & \colhead{Defining equation}}

  \startdata
  \sidehead{(Pre-impact) CSM parameters:}
    $\rhocsm$    & mass density          & $\densu$ &         \\  
    $n_\csm$     & particle density      & ${\rm cm^{-3}}$ & \\
    $m_p$        & proton mass                           & g & $1.673\times10^{-24}$~g \\
    $\mu$        & mean molecular weight &  & $\rhocsm = \mu m_p n_\csm$ \\
    $N_\csm$          & column density   & ${\rm g~cm^{-2}}$ & Equation \ref{eqn:N} \\
    $R_{\rm in}$ & inner radius               & cm     & HNKEq 5 \\
    $f_R$        & fractional width           &        & $\Delta R/R_{\rm in}$ \\  
    $M_\csm$     & mass                       & $\Msun$ & Equation \ref{eqn:M} \\
  \sidehead{light-curve parameters:}
    $\timp$      & time of ejecta impact with CSM & days   &         \\
    $t_p$        & time of radio peak; time that forward shock overruns CSM & days & HNKEq 7 \\
    $\nu$        & photon frequency               & Hz     & \\
    $\epsilon_B$ & ratio of magnetic field energy density to gas energy density &  & \\
    $p$          & power-law slope of relativistic electron density &  & HNKEq 25 \\
    $\Lp$        & specific luminosity at radio peak & ${\rm erg~s^{-1}~Hz^{-1}}$ & HNKEq 11 
  \enddata

  \tablecomments{HNKEq refers to the equation number in HNK16.}
\end{deluxetable*}

The important hydrodynamic conclusion of the work was a simple equation for
the time that the forward shock will reach the outer edge of the CSM shell, 
which we sometimes refer to as the ``end'' of interaction.
As in HNK16 Equation~5, the inner CSM radius ($R_{\rm in}$), 
impact time ($\timp$), and CSM density ($\rhocsm$) 
are related through
\begin{eqnarray}
  R_{\rm in} \propto \timp^{0.7}~\rho_\csm^{-0.1} \quad .
  \label{eqn:Rin}
\end{eqnarray}
Because the shells have constant density, the mass of a shell is determined by
$R_{\rm in}$, $\rhocsm$, and 
the CSM fractional width ($f_R \equiv \Delta R/R_{\rm in}$) simply by
\begin{eqnarray}
  M_\csm = \frac{4\pi}{3} \rhocsm R_{\rm in}^3 \left[ (1+f_R)^3 - 1 \right] \quad .
  \label{eqn:M}
\end{eqnarray}
Given in HNK16 Equation~7, 
the time that the forward shock reaches the outer edge of the CSM shell
(the time of the radio peak luminosity, $t_p$)
is related to $\timp$ and $f_R$ by
\begin{eqnarray}
  t_p \propto \timp~(1+f_R)^{1.28} ~.
  \label{eqn:tp}
\end{eqnarray}
Motivated by nova shells, in HNK16 $f_R = [0.1,1]$, but we ensured through
simulation of thicker shells that this relation holds at least up to
$f_R = 7$.
Using this equation is only appropriate for adiabatic shocks, which we 
assume is appropriate for our low-density shells;
HNK16 Figure~1 shows where cooling is expected to become important.
The relation is only weakly dependent on the exact velocity at
which the ejecta is truncated, as explored in HNK16 
(Figure~7).

The important radiation conclusion was a parameterization for the 
radio synchrotron light-curves.
The time that the forward shock reaches the outer edge of the CSM shell 
is the time of radio peak luminosity.
The optically thin peak specific luminosity 
($\Lp$, units ${\rm erg~s^{-1}~Hz^{-1}}$) is related to 
$\rhocsm$, $R_{\rm in}$, $f_R$, 
and frequency ($\nu$) in HNK16 Equation~11,
\begin{eqnarray}
  \Lp \propto \nu^{-1}~\rhocsm^{8/7}~R_{\rm in}^{3/7}~ \left[ 1 - (1+f_R)^{-9/7} \right] ~,
  \label{eqn:Lp}
\end{eqnarray}
where the exponent 9/7 is an approximation of the fit value 1.28 to elucidate the relative 
importance of each factor.
As in HNK16, we use $\epsilon_B = 0.1$ as the fraction of energy density 
in the magnetic field compared to the gas thermal energy density.
HNK16 Figure~1 shows which shells are expected to be optically
thin at all times.
After peak, the radio emission declines rapidly because 
the CSM shell, heated and accelerated by the shock, quickly 
rarefies into the near-vacuum that lies outside its outer edge
(HNK16 Figure~2). 
Thus, even a light-curve that was not optically thin to synchrotron
self-absorption while the shock was inside the shell is likely to 
become optically thin shortly after peak. 
HNK16 provides a parameterization for the light-curve decline
considering only emission from the shocked CSM
(HNK16 Equation~12), which describes the decline from 
$\lumL_p$ to $10^{-3}\,\lumL_p$ using only $\timp,~f_R$, and $\Lp$.
We cannot comment on early impacts with thin shells because they would have 
declined below $10^{-3}\,\Lp$ by the time of observation and are therefore
not captured by this parameterization.
A parameterization including the reverse shock emission is 
given in HNK16 Equation~13. 
Examples of radio light-curves are shown in HNK16 Figures~3 and 4,
with Figure~3 showing the light curve including reverse shock
emission.

These models were developed with thin nova shells in mind; in such a thin
shell, the approximation of constant density is appropriate. 
How accurately constant-density shell describes a more extended CSM like that 
of PTF11kx \citep{Silverman+13b,GH+17} is unknown.

\section{Observations}
\label{sec:obs}

\begin{figure}
\centering
  \includegraphics[width=3in]{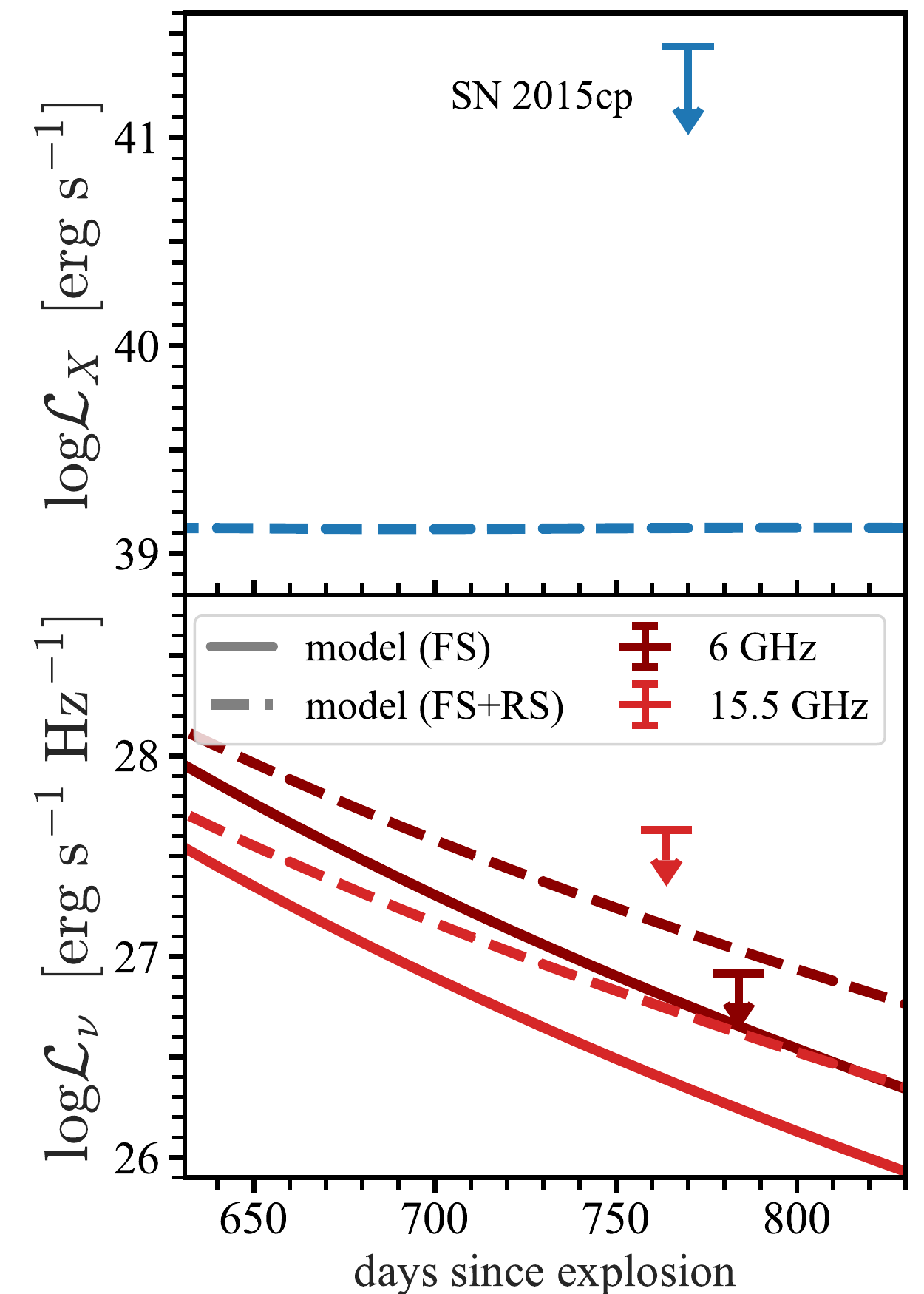}
  \caption{Observations of SN~2015cp (arrows) in 
          X-ray (top panel; 0.5--8 keV {\it Swift}) 
          and radio (bottom panel; AMI at 15.5 GHz, bright red; 
          VLA at 6 GHz, dark red) compared to a model with  
           $\rho_\csm = 10^{-19}~\densu$, $f_R = 4$, and $\timp=50~\days$
           created for PTF11kx both including (solid) and 
           excluding (dashed) contribution from the reverse shock.
           }
  \label{fig:obs}
\end{figure}

Radio and X-ray modeling of the nearest analog to SN~2015cp, PTF11kx, 
indicated that a PTF11kx-like event at the distance of SN~2015cp 
($\sim 170~{\rm Mpc}$) would be visible in the radio and X-rays, 
though fading (Figure \ref{fig:obs}).
The CSM parameters for these models come from \citet{GH+17},
and the model shown represents the lowest-density possibility for
PTF11kx.
Synchrotron and bremsstrahlung emission were calculated as in
that work and HNK16, both including (dashed lines) and excluding 
(solid lines) emission from the shocked ejecta \citep[Gaunt factors from][]{vanHoof+14}
Unfortunately, no X-ray or radio observations of PTF11kx years after maximum light 
are known to the authors for a direct comparison. 
The light curves demonstrate that AMI and VLA observations had a chance of
detecting thick, low-density shells even hundreds of days after 
interaction has ended. 
Though not shown here, X-ray observations were promising for denser shells.
We therefore pursued observations with the Arcminute Microkelvin Imager (AMI), 
the Jansky Very Large Array (VLA), 
and the Neil Gehrels {\it Swift} Observatory.

SN~2015cp was observed by the AMI \citep{Zwart+08,Hickish+18} on 04-Dec-2017/19:30:17.7 UT (MJD 58091.82) with a total integration time of 4.96~hr. The AMI-LA is equipped with a digital correlator \citep{Hickish+18} with a central frequency of $15.5\,\textrm{GHz}$ and a $5\,\textrm{GHz}$ bandwidth spread across 4096 channels. The data were calibrated using the custom reduction pipeline \textsc{reduce\_dc} \citep[see, e.g.,][]{Perrott+13}, which also performs flagging for antenna shadowing, instrumentation errors, and radio frequency interference (RFI). At this stage the data were binned into 8 frequency channels and imported into the Common Astronomy Software Applications (CASA) package, where additional RFI flagging and then cleaning were performed to produce an image, which contained a single unresolved source. Fitting this source with the CASA task IMFIT gives its J2000 location as $\alpha = 03^h09^m13.18(2)^s$, $\delta = +27^\circ 34' 16.8(5)''$ with a flux density of $1.24\pm0.08\,\textrm{mJy}$. Given that this position is $\sim3'$ from the phase center and the synthesised beam for this observation has major and minor axes of $\sim40''$ and $30''$ (respectively), the object is probably not SN~2015cp. This conclusion is further supported by the position being consistent with a known source in the NVSS archive \citep{Condon+98}. The root-mean square (RMS) at the phase center of the AMI-LA image is $\sim30\,\mu\textrm{Jy}$, so we set a 3$\sigma$ upper limit on the $15.5\,\textrm{GHz}$ radio emission from SN~2015cp of $\sim90\,\mu\textrm{Jy}$.

On UT 24-Dec-2017, we observed SN~2015cp with the VLA (under program 17B-434, PI Horesh) at a central frequency of 
6 GHz (C-band) and undertaken in the B configuration. 
We used J0329+2756 and 3C~138 for phase and flux calibration, 
respectively, and reduced the data using standard CASA calibration and imaging routines. 
The observation resulted in a null detection with a 5.8 $\mu$Jy RMS, 
corresponding to a $3\sigma$ upper limit of 17.4 $\mu$Jy.

From UT 8-Dec-2017 through 9-Dec-2017, we observed SN~2015cp with the UVOT and XRT instruments on {\it Swift}.  No source is detected in a 4.9~ks UVOT exposure using the UVW1 filter, nor are any counts detected near the source position in the 6.1~ks (livetime, 0.5--8 keV) XRT exposure.  Using the {\it Swift} UVOT zeropoints\footnote{https://swift.gsfc.nasa.gov/analysis/uvot\_digest/zeropts.html}, we derive a 3$\sigma$ limiting magnitude of 23 in the AB system.  Assuming only absorption by the Galaxy in the direction of the SN \citep[with $N_{\rm H} = 1.1 \times 10^{21}$ cm$^{-2}$;][]{Kalberla+05} and assuming a spectrum with photon index $\Gamma=2$, we derive a 3$\sigma$ limiting X-ray flux of $4.1 \times 10^{-14}$ erg cm$^{-2}$ s$^{-1}$ (0.5--8 keV).

\begin{deluxetable}{llll}
  \tablecolumns{4}
  \tablewidth{\columnwidth}
  \tablecaption{SN~2015cp Radio and X-ray Observations
                \label{tab:obs}}
  \tablehead{\colhead{UT Date} & \colhead{Instrument} & \colhead{Observed Frequency} & \colhead{3$\sigma$  limit}}

  \startdata
    2017-12-04   & AMI-LA      & 15.5 GHz   & 90 $\mu$Jy \\  
    2017-12-24   & VLA         & 6 GHz      & 17.4 $\mu$Jy \\
    2017-12-08   & Swift-UVOT  & 4.8 eV     & 2.3 $\mu$Jy \\
    2017-12-08   & Swift-XRT   & 1 keV      & 6.1 nJy 
  \enddata
\end{deluxetable}

\section{Limits on the CSM of SN~2015cp}
\label{sec:analysis}

In this section we use the constant-density, finite-extent 
CSM radio light-curve models of HNK16 to limit the CSM 
properties given our radio nondetections.
The direct observational considerations in this analysis are that
(1) impact occurred at $> 60~$ days,
(2) our {\it earliest} radio limit is at 764 days, and
(3) our {\it deepest} radio limit is at 784 days.
In \S\ref{ssec:cases} we will consider the NUV and optical observations
to argue that the bulk of the CSM had been swept over by the time of 
the NUV observation, and define two scenarios that we consider likely for
the time of radio peak (Cases 1 and 2).
In \S\ref{ssec:csmlims} we translate the radio upper limits into 
CSM mass limits, from which we see the importance of regular monitoring
of interaction candidates and of rapid radio follow-up observations.
Finally, we investigate other properties of the maximum-mass CSM shells
allowed by our radio nondetections and show that their column densities
are high enough that pre-impact spectra could show narrow absorption 
features as in PTF11kx; this, combined with the lack of absorption features
in the pre-impact spectrum, indicates that the CSM is probably low density
as assumed.

\subsection{Constraining the Duration of Interaction}
\label{ssec:cases}

In \S\ref{ssec:csmlims} we will see that constraining the time at which interaction ends 
($t_p$; the time the forward shock reaches the edge of the shell) 
or the time of impact ($\timp$) increases the utility of the radio data. 
To the purpose of constraining the interaction timescale, 
we compare SN~2015cp to its nearest analog, PTF11kx,
using the metric that indicated the interaction history of that object: 
the integrated luminosity of the broad \Hal\ line ($\Lhal$) from \citet{Silverman+13b}.

\begin{figure}
\centering
  \includegraphics[width=3.2in]{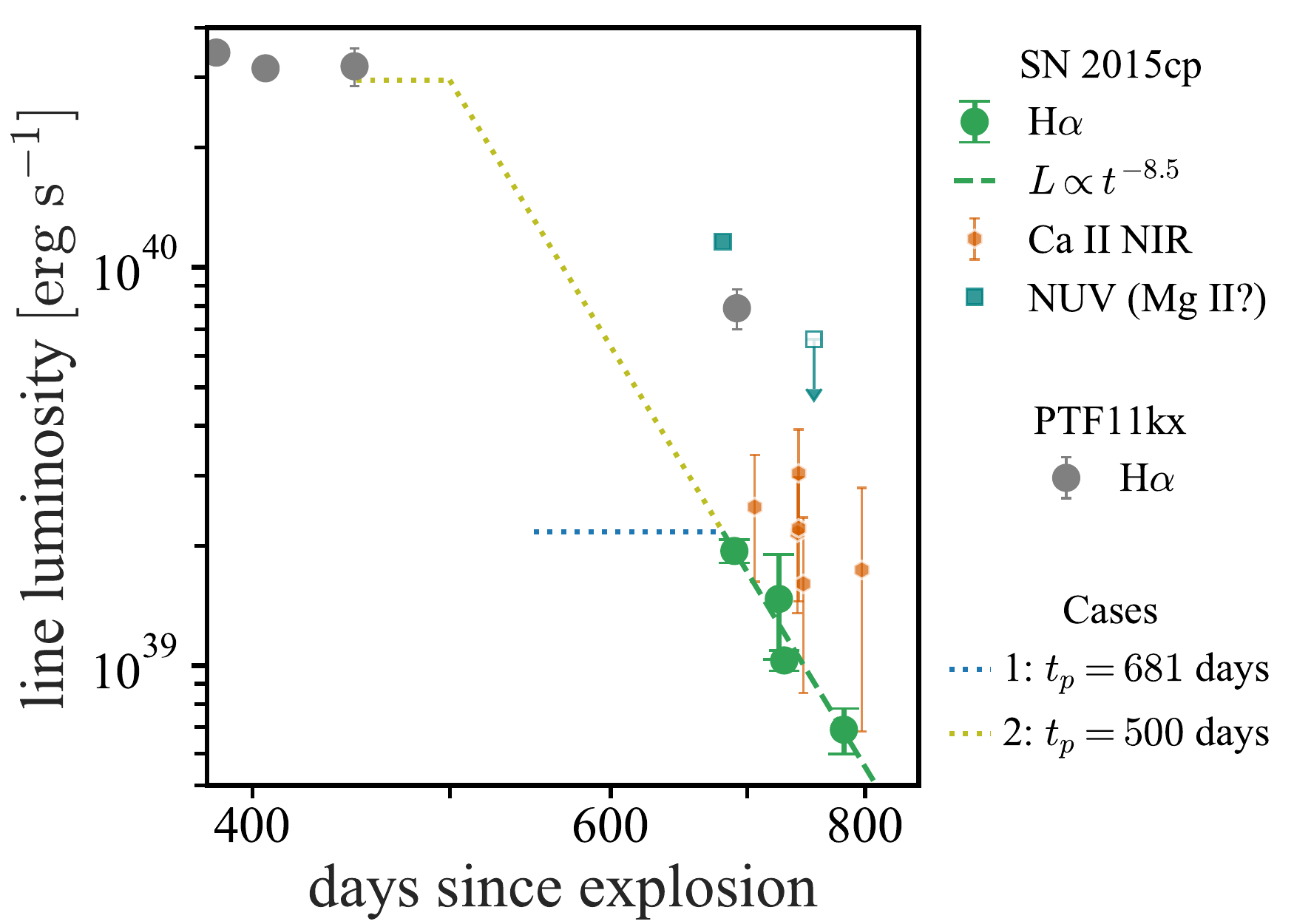}
  \caption{The evolution of the broad \Hal\ emission-line luminosity
  	       constrains the duration of interaction for SN~2015cp (green circles), 
           as for PTF11kx \citep[][grey circles]{Silverman+13b}.
           The steep decline ($\lumL \propto t^{-8.5}$;green dashed line)
           indicates interaction was over by the time of 
           the NUV detection at 681 days.
	       The integrated \CaII\ near-infrared emission (orange hexagons)
           and the NUV data that may have been line emission from \MgII\ 
           (teal squares) are consistent with the \Hal\ decline rate. 
           In this work, we consider two scenarios (dotted):
           Case 1 (blue), that SN 2015cp is intrinsically fainter
           than PTF11kx and we discovered SN~2015cp just as interaction 
           ended ($t_p = 681~\days$);
           and Case 2 (yellow), that SN~2015cp had the same \Hal\ luminosity 
           as PTF11kx and its interaction ended at $t_p = 500~\days$.
          }
  \label{fig:decLINE}
\end{figure}

In Figure~\ref{fig:decLINE} we show $\Lhal$ for PTF11kx and SN~2015cp.
Note that the point with large error bars is a combination of three measurements,
with uncertainties reflecting systematic and statistical errors.
For SN~2015cp we also include the line luminosity of the \CaII\ near-infrared (NIR)
triplet and the NUV observations presented and discussed by \citet{Graham+18} 
which may be an \MgII\ line.
The \CaII\ and NUV signals have decline rates consistent with that of $\Lhal$.
No SN~Ia features are seen in the spectrum; 
all of these lines are powered by interaction.

As seen in Figure~\ref{fig:decLINE}, PTF11kx and SN~2015cp have very different 
spectroscopic coverage: 
the former has observations throughout the years, 
whereas the latter sampling is of the decline only but at 
comparatively high cadence.
For PTF11kx, we know $\Lhal$ increased until 285 days, 
plateaued for the next 160 days, 
and started to decline sometime between 450 and 695 days. 
The decline was interpreted as the time at which the shock had
swept over the majority of the CSM such that the reservoir of shocked, 
cooling gas was depleted \citep{Silverman+13},
constraining $450 \leq t_p \leq 690~\days$ yet leaving the $\Lhal$
decline rate highly uncertain.
(We note that this interpretation is not based on
detailed hydrodynamic and radiation transport calculations, 
so there is an additional but currently unquantified level of 
uncertainty in this interpretation that requires sophisticated
modeling to understand, which is far beyond the scope of this work.)
In contrast, SN~2015cp has no spectra between its classification 
spectrum ($\sim 60~\days$) and the X-Shooter spectrum ($706~\days$), 
so its impact time and peak times are poorly constrained.
But its decline is well observed by the follow-up campaign.
Over the observed 100 days, SN~2015cp has $\Lhal \propto t^{-8.5}$
(from an error-weighted fit to the data performed with {\tt scipy}).

The fortuitous observations of both SNe with optical spectra at 
$\sim60$ and 700 days allows us at least to say that these 
SNe are not twins. SN~2015cp is an order of magnitude fainter in \Hal\ at late
times, and began interacting with its CSM later --- the classification spectrum
was of sufficient quality to detect, at 3$\sigma$ confidence, a narrow \Hal\ line an order of magnitude less 
luminous than was observed in PTF11kx \citep{Graham+18}.

If we knew the peak \Hal\ luminosity reached by SN~2015cp, we would be able to
constrain the time interaction ended.
Since we do not have these data, we consider two cases representing
different conservative assumptions, illustrated in Figure \ref{fig:decLINE}.
\begin{itemize}
  \item Case 1 assumes SN~2015cp was brightest in \Hal\ at NUV discovery, i.e., $t_p = 681~\days$.
		This would not be as serendipitous as it may, at first, seem: 
		detection favors the bright, and the 
		\Hal\ luminosity was likely highest just before interaction ended --- when the 
		reservoir of cooling shocked hydrogen is greatest. 
  \item Case 2 assumes the observed $\Lhal$ decline rate is constant 
        {\it and} that SN~2015cp had the same maximum $\Lhal$ as PTF11kx, 
        resulting in $t_p = 500~\days$.
        Furthermore, if PTF11kx and SN~2015cp had the same CSM density, then, like PTF11kx, 
		SN~2015cp would have needed 282 days to reach the plateau luminosity, 
		so for this case $\timp \leq 218~\days$. 
        Furthermore, this comparison implies that for PTF11kx $t_{p,{\rm 11kx}}=588~\days$ 
        (if it follows the same decline rate) so $f_{R,{\rm 11kx}} = 5.9$.
\end{itemize}
Case 1 is our favored scenario because it has fewer assumptions:
Case 2 relies on both extrapolation and a comparison to PTF11kx that may not 
be appropriate. 

Finally we note that the limits obtained in Cases 1 and 2 bound what would be obtained 
from assuming the $t^{8.5}$ decline rate and any $500 \leq t_p \leq 681$.

\subsection{Mass Limits from Radio Data}
\label{ssec:csmlims}

Using these parameterized light curves described in \S\ref{sec:mod_desc},
we can determine the maximum CSM mass allowed by our VLA and AMI nondetections, with
particular consideration given to the two cases put forth in the previous section.

The methodology is straightforward. 
We create a grid of models varying $\timp$ and $f_R$ such that there are a variety of
peak times (i.e., a variety of delay times between radio peak and AMI/VLA observation).
For each model light curve we know the maximum $\lumL_p$ allowed by the AMI/VLA
limit, which translates into a maximum allowed density.
From $\timp$, $f_R$, and the maximum allowed density we then know the maximum
allowed mass (see Equations~\ref{eqn:Rin} and \ref{eqn:M}).

From observations, $\timp\geq60~\days$ and $t_p\leq681~\days$.
In line with HNK16, we only consider $f_R\geq0.1$.
These criteria set our explored parameter space to 
$t_{\rm imp}\in[60,616]~{\rm days}$ and $f_R\in[0.1,5.74]$. 
Cases 1 and 2 fix $t_p$ to a particular value and represent a specific contour
in $\timp$--$f_R$ space 
(and in Case 2, $\timp\leq208~\days$, i.e., $f_R\geq1.02$; Equation~\ref{eqn:tp}).

\begin{figure}
\centering
  \includegraphics[width=3.2in]{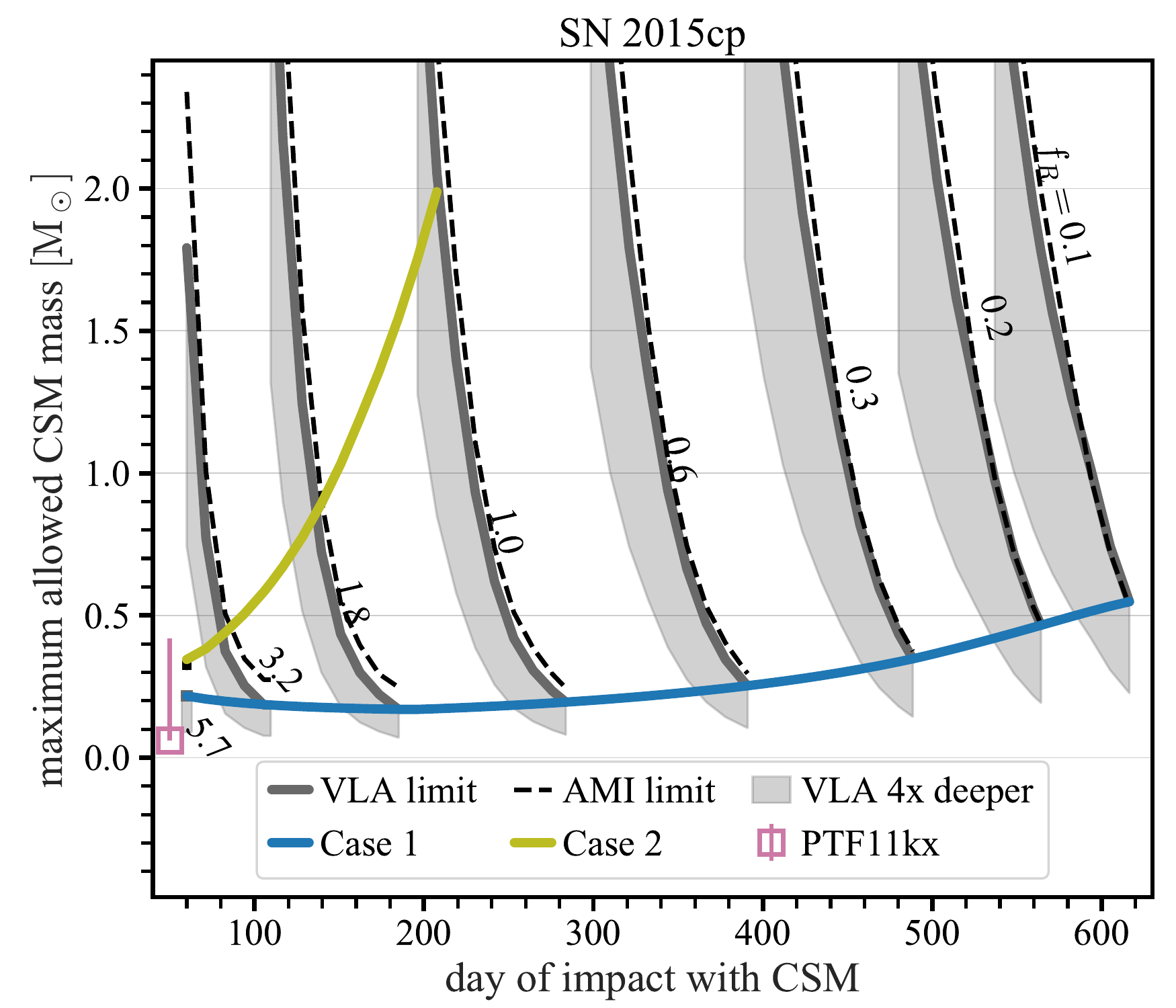}
  \caption{Upper limits on the CSM mass of SN~2015cp as a function
  of the impact time ($\timp$) and shell width ($f_R$),
  assuming a finite-extent, constant-density CSM shell.
  PTF11kx is shown for comparison (pink open square)
  for $f_R=4-6$ (pink line).
  Constraints from the 769 day VLA 6 GHz nondetection (grey line)
  and the 750 day AMI 15.5 GHz nondetection (dashed black line) are similar,
  highlighting the power of prompt observations.
  For reference, we illustrate how the constraints would 
  change if the VLA nondetection were four times deeper (grey shading).
  Cases 1 (blue line) and 2 (yellow line) assume the time of radio peak to 
  681 and 500 days, respectively, and show the importance of constraining
  $t_p$.
  The CSM mass is constrained to
  $M_\csm < 0.5~M_\odot$ for Case 1 (and $<2~M_\odot$ for Case 2) 
  {\it despite large uncertainties in the CSM location and extent}.
  }
  \label{fig:csm_mass}
\end{figure}

The mass limits resulting from this analysis are shown in Figure \ref{fig:csm_mass}. 
It is important to acknowledge that the mass constraints from the 
VLA data and the AMI data are nearly the same, highlighting the power 
of rapid follow-up observations for these steeply declining light curves.
The VLA limit is eleven times deeper than that of AMI, after accounting for the 
spectral energy distribution (SED) and the observed frequency, 
but the AMI limit is as effective simply by virtue of
occurring three weeks earlier.
For reference, we also present how the maximum allowed mass would decrease if the VLA 
nondetection were four times deeper, to simulate a longer exposure time or a closer
object and to indicate that these are upper limits. 

Given an $f_R$, there is a maximum possible impact time that 
satisfies the requirement $t_p < 681~\days$, setting the latest $\timp$ 
for each $f_R$ curve. The thickest possible shell has only one valid impact time, 
while thinner shells span more of the domain.
For curves thinner than $f_R \approx 1$, the shaded regions
also appear to have an earliest possible impact time, but this is simply due to
the limitations of the parameterized light curves (\S\ref{sec:mod_desc}). 
The steep slope of each $f_R$ curve reflects the steep decline of the light curves.

The degeneracy between $f_R$ and $\timp$, as well as the sensitive dependence of
the mass limit on $\timp$, are obvious in this plot and motivated the search
for additional constraints on the duration of interaction described in \S\ref{ssec:cases}.
We consider two cases: Case~1, the radio light curves peak at $t_p = 681~\days$;
and Case~2, $t_p = 500~\days$. 
A fixed $t_p$ gives a relation between $f_R$ and $\timp$
rather than a constraint on either parameter alone (Equation \ref{eqn:tp})
as demonstrated by the curves in Figure~\ref{fig:csm_mass}.
In fact, $t_p$ constrains the CSM mass very well almost independent of the 
inaccuracy in $\timp$ and $f_R$. For this reason, we identify it as the key parameter
to constrain in future efforts to characterize SNe~Ia;n.

The VLA nondetection constrains the 
CSM mass to $M_\csm < 0.5~M_\odot$ for Case~1 and $M_\csm < 2~M_\odot$ for Case~2.
Particularly in Case 1, these limits are similar to the mass measured for 
PTF11kx \citep{GH+17}, as shown. We show both the reported $M_{\rm csm,11kx}=0.06~M_\odot$ 
as well as a higher estimate of $M_{\rm csm,11kx} = 0.42~\Msun$
that results from assuming the four-times-higher column density reported
by the authors as well as the result from this work that the CSM may have been thicker 
than previously assumed (\S \ref{ssec:cases}).

\subsection{The Constraining Power of Pre-Impact Spectra}
\label{ssec:N}

In this section, we translate the VLA upper limit into limits on the CSM
column density ($N_\csm$) and its constituent factors: $f_R$, $R_{\rm in}$, and 
$\rhocsm$.
For PTF11kx, the saturated narrow absorption lines in the pre-impact 
optical spectra alerted observers to the unique nature of the event. 
\citet{Borkowski+09} identify the narrow Na D features as a good metric for 
CSM shell properties when considering shells at $>10^{17}~{\rm cm}$, and 
we expect the same is true for nearer cases.
This analysis is relevant for future observations rather than SN~2015cp itself,
whose only pre-impact spectrum is too late to have revealed such features.
The measurements for PTF11kx referenced in this analysis come from \citet{GH+17}.

The constituents of the column density are $f_R$, $R_{\rm in}$, and $\rhocsm$:
\begin{equation}
	N_\csm = n_\csm f_R R_{\rm in} = \rhocsm f_R R_{\rm in}/(\mu m_p) \quad .
    \label{eqn:N}
\end{equation}
For this analysis we have assumed the mean molecular weight $\mu = 1.33$. 
In the above equation, $n_\csm$ is the particle density and $m_p$ the proton mass.

\begin{figure}
\centering
  \includegraphics[width=3.2in]{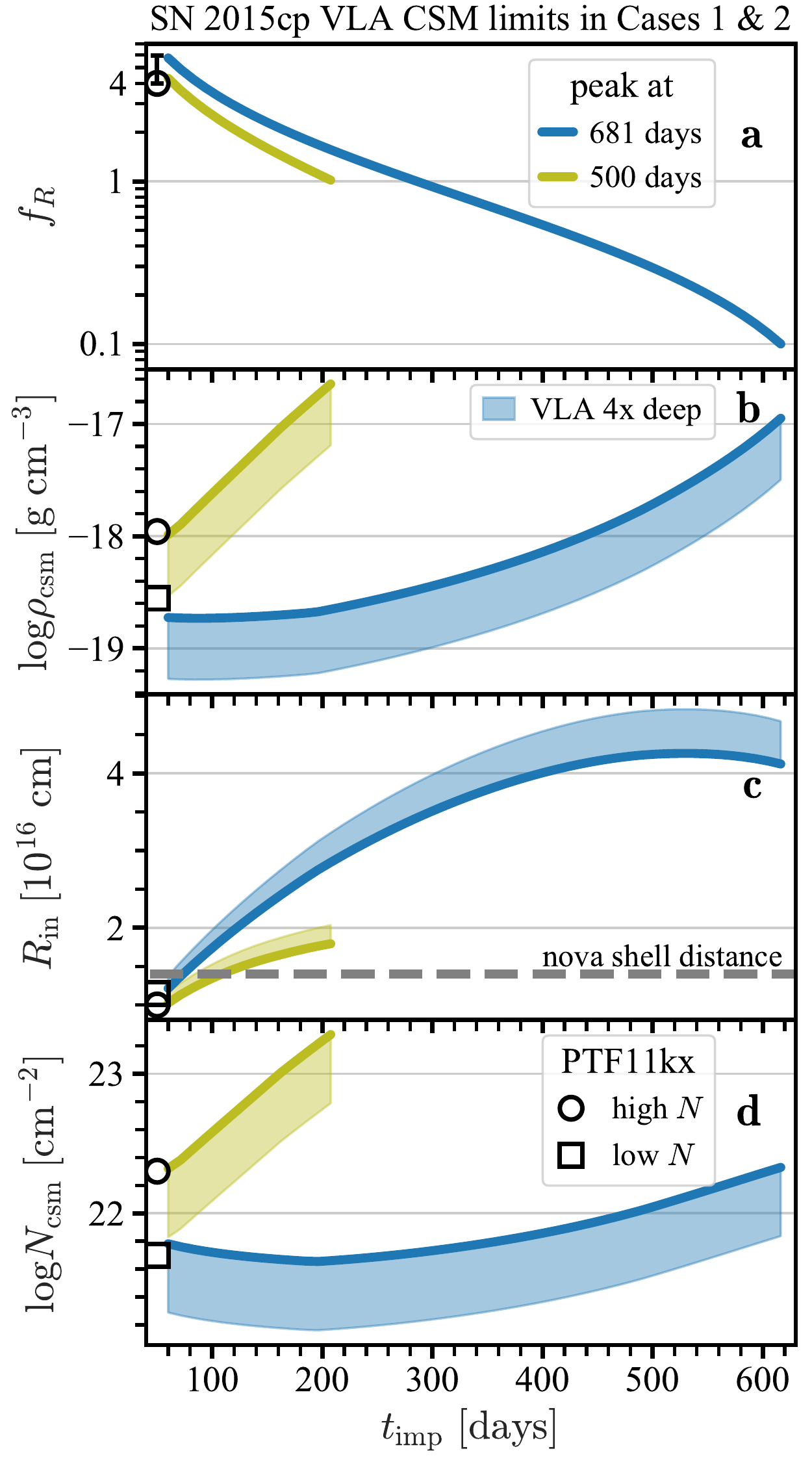}
  \caption{Constraints on CSM properties for Case 1 (blue; radio peak at 681 days) and 
           Case 2 (yellow; peak at 500 days), with a comparison to PTF11kx (open markers).
           Shaded bands represent the analysis with a nondetection up to 
           four times deeper to illustrate limits. Legends apply to all panels.
           The panels are
  		   (a) fractional width,
           (b) mass density,
           (c) inner radius,
           and 
           (d) column density.
           Our limits on the inner radius are similar to the expected location of 
           nova shells.
	       Panel {\it (d)} shows that a measurement of column density from an 
           optical spectrum near the SN $B$-band maximum could be leveraged, 
           e.g., $3\times10^{22}~\mathrm{cm^{-2}}$ would rule out Case~1.
           }
  \label{fig:csm_props}
\end{figure}

In Figure~\ref{fig:csm_props} we show these constraints for Cases~1 and 2
again as a function of $\timp$. 
As in Figure~\ref{fig:csm_mass}, shaded bands represent a simulated VLA nondetection 
up to four times deeper than the actual nondetection. We also show the value of
each parameter for PTF11kx as derived by \citet{GH+17}, noting both the higher
and lower reported values for the column density. 
The equations provided in \S\ref{sec:mod_desc} are a helpful reference when 
interpreting the behavior of the curves in panels {\it (a)--(c)}.

Panel {\it (a)} simply illustrates the relationship between $f_R$ and $\timp$ 
for a fixed $t_p$ and is independent of the VLA limit. 
Earlier $\timp$ values require wider shells to peak at a given time.
The error bar on the PTF11kx point indicates the fractional width
$f_{R,{\rm 11kx}} = 5.9$ 
that would be implied if interaction ended at 588 days 
(\S\ref{ssec:cases}).

Panel {\it (b)} shows upper limits on CSM density.
The Case 2 limit is higher (weaker) than the Case 1 limit because
the light curve has had longer to decline, so higher peak luminosities and 
thus higher densities are permissible. 
Later impact times also allow higher $\rhocsm$ (for a given $t_p$)
because thinner shells have a lower peak luminosity and faster decline
rate than thicker shells.
For PTF11kx this value is derived using measurements of $f_R$, $\timp$, and $N_\csm$.

In panel {\it (c)} we plot the limit on $R_{\rm in}$, which is actually a lower
limit. While generally the behavior of this curve is intuitive 
(one would expect a later impact time to imply a more distant shell),
the weak dependence of $R_{\rm in}$ on $\rhocsm$ actually requires that 
higher-density shells be nearer to the SN~for interaction to begin at a given
time, resulting in a maximum $R_{\rm in}$. 
The dependence on $\rhocsm$ is also why a deeper VLA limit (shaded band),
constraining the CSM to lower densities, would imply a higher $R_{\rm in}$. 
The value for PTF11kx comes from $\timp$ and assuming a maximum ejecta speed,
but is consistent with the HNK16 value given its $\timp$ and $\rhocsm$.

Finally, in panel {\it (d)} we use Equation \ref{eqn:N} to translate our 
radio nondetections into an upper limit on $N_\csm$. 
As with the mass constraint previously, fixing $t_p$ results in limits on $N_\csm$ 
that are almost independent of $\timp$, particularly for Case 1. 

While $N_\csm$ is not single-valued, we see that a reliable measurement of $N_\csm$ 
from a spectrum near maximum $B$-band brightness could
constrain $t_p$ by making Case 1 (i.e.,\ late values of $t_p$) less likely.
The Ca K absorption line in PTF11kx that was used to measure $N_\csm$ was strong
up to at least 20 days after maximum light, so this metric does not necessarily
require very early-time spectra (but does benefit from high-resolution spectra).

\section{Conclusions and Discussion}
\label{sec:conc}

SN~2015cp is the third case of an SN~Ia interacting with CSM located 
$\sim 10^{16}~{\rm cm}$ from the progenitor system, what we call an
SN~Ia;n. 
The ``X;n'' label designates a supernova that initially appears normal, 
but weeks to months after SN peak is dominated by interaction signatures. 
Here we will summarize our radio and X-ray nondetections
and the constraints on the circumstellar environment of SN~2015cp obtained via 
the models of \citet[][``HNK16'']{HNK16}.
We will then put these limits in context of SN~Ia progenitor theory and other known
interacting SNe~Ia.
Finally, we will highlight the need for a systematic search for SNe~Ia;n and
make suggestions based on the lessons learned from our analysis of both PTF11kx and 
SN~2015cp.

This work relies on the hydrodynamic modeling and radiation calculations
of HNK16, which we summarized briefly in \S\ref{sec:mod_desc}.
These models address the scenario of an adiabatic shock propagating through 
a constant-density CSM shell with a distinct inner and outer edge.
When the forward shock reaches the edge of the CSM, we say interaction
has ``ended'' because the CSM is no longer gaining thermal energy; this
is the time of peak radio luminosity.
In HNK16, a simple description of the relevant hydrodynamic timescales and
the optically thin radio synchrotron light curves was derived. 

As described in \S\ref{sec:obs} and shown in Figure~\ref{fig:obs}, 
at a distance of 167 Mpc, SN~2015cp was a good candidate for radio and X-ray 
follow-up observations once interaction was discovered. 
Models of its nearest analog, PTF11kx, suggested SN~2015cp could still
be visible even if (like PTF11kx) interaction had ceased --- 
that interaction had ended for SN~2015cp was unknown at the time of the observations.
Therefore, we observed this target with the VLA, AMI, and \Swift;
but all observations resulted in nondetections.

Nevertheless, the radio upper limits can be interpreted in the 
framework of \citet{HNK16} to give limits on the CSM properties.
The extent of the CSM is a key parameter in these models, and can
be determined if the time of impact and time that the forward
shock sweeps over the bulk of the CSM are known. In the case of
SN~2015cp, we can only place limits on these times (\S\ref{ssec:cases}).
For both SN~2015cp and PTF11kx, the limits on the duration of 
interaction come from the evolution of the H$\alpha$ luminosity.
The decline of both the optical and NUV emission suggests that
the shock was no longer energizing CSM by the time of our observations.
We assess two scenarios, which can be thought of as late and early
limits on the time interaction ended: 
Case 1, in which the radio light curve
peaked at the time of NUV discovery (681 days); 
and Case 2, in which the radio light curve peaked at 500 days,
under the assumption that the \Hal\ luminosity of SN~2015cp was
the same as that of PTF11kx and a constant decline rate (Figure~\ref{fig:decLINE}).
Without these assumptions there is too large an uncertainty in $\timp$--$f_R$
space to draw conclusions (Figure~\ref{fig:csm_mass}).

In \S\ref{ssec:csmlims}, we find that the CSM has a mass 
$M_\csm < 0.5~M_\odot$ for Case 1. 
Case 2 is less constraining because it implies our observations occur 
longer after peak radio brightness than in Case 1; 
in Case 2, $M_\csm < 2~M_\odot$ (Figure \ref{fig:csm_mass}).
These limits are near to the inferred CSM mass of PTF11kx, 
but far higher than the estimated mass from a single nova shell eruption,
which for the symbiotic recurrent nova RS Ophiuchi was observed to be 
$\sim 10^{-6}~\Msun$~\citep{OBrien+92}.

We also explored the limits on CSM extent, density, inner radius,
and column density that can be derived from the radio observations
(\S\ref{ssec:N}; Figure \ref{fig:csm_props}).
We find that in Case 1 the CSM of SN~2015cp had, at most, 
the same column density as PTF11kx. 
For Case 2, higher column densities are allowed.
The lack of any narrow absorption features in the pre-impact spectrum
supports the idea that SN~2015cp had a lower column density than 
PTF11kx, which in turn implies a lower density
($\sim 10^{-19}~{\rm g~cm^{-3}}$), though it may be the case that the
spectrum was too late time to see the feature.

\begin{figure*}
\centering
  \includegraphics[width=3.5in]{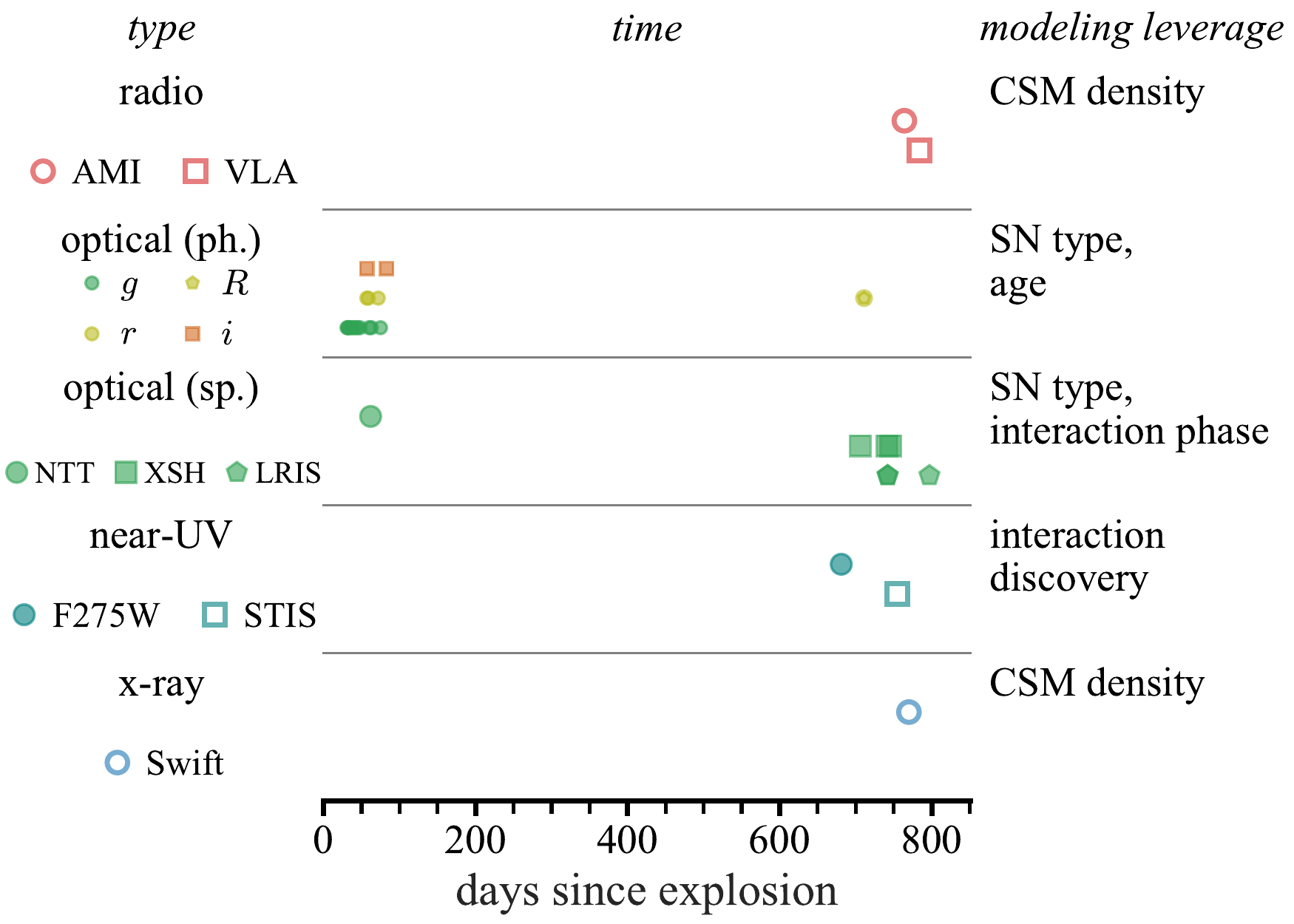}
  \caption{A summary of the panchromatic data available for SN~2015cp
           and how they are used in our analysis.
           Optical observations are split by whether they are 
           photometric imaging (``ph.'') or spectroscopy (``sp.'').
           Legends indicate the instrument or filter used, and 
           nondetections are open markers.
           The two-year gap between the initial discovery and classification 
           of this supernova and the discovery of its interaction creates large
           uncertainties in its CSM properties.
           }
  \label{fig:summ}
\end{figure*}

Figure \ref{fig:summ} summarizes how the various observations of 
SN~2015cp have been employed to constrain the CSM properties.

As with promptly-interacting SNe~Ia,
SNe~Ia;n are associated with the ``91T-like'' subclass.
Here we define this as spectroscopically most similar to SN~1991T {\it or}
SN~1999aa near maximum light, with the light-curve similarities being a secondary
classification metric when a near-maximum spectrum is lacking.
The physical connection between these three groups is undetermined.
The simplest hypothesis to explain the similarity of the spectroscopic 
SN features is that the groups share the same progenitor system.
But, particularly for the prompt interaction cases, 
which lack a noninteracting phase for studying the SN alone, 
some debate whether the progenitor is a high mass star \citep[e.g.,][]{Inserra+16}. 
\citet{Silverman+13} also noted that PTF11kx differed from promptly-interacting 
SNe~Ia-CSM.
Further studies both observational and theoretical of SNe~Ia;n are needed 
to establish the connection between 91T-like SNe Ia, SNe~Ia;n, and promptly-interacting 
SNe~Ia-CSM for certain.

If the systems do have a common progenitor, 
eruptive mass loss episodes could naturally bridge 
noninteracting, late-interacting, and promptly-interacting SNe~Ia.
An eruptive mass loss episode can quickly (within years) 
sweep extended CSM into to a shell at distance of $\sim 10^{16}~{\rm cm}$ 
\citep{MooreBildsten12} which would then mix with the interstellar medium 
within $\sim~10^{6}~{\rm yr}$. 
This would explain why some 91T-like SNe Ia have clean environments (like SN~1991T)
while others interact with CSM at $\sim 10^{16}-10^{17}~{\rm cm}$.
Furthermore, multiple eruptions will result in
CSM shells collected near the same radius such that the CSM mass observed in the
interaction need not represent that of a single eruption (and swept up material). 
Prompt interaction would be observed if the SN occurred while extended material
was still in the progenitor system, i.e., before an eruption.
Thus, the three observed groups are natural in an eruptive-mass-loss scenario.

However, even determining that eruptions form the CSM does not by itself reveal 
the nature of the progenitor system, as there are a variety of mechanisms for 
rapid mass loss that evacuate an inner cavity.
Novae and symbiotic novae are the canonical example, and tied to the single-degenerate
progenitor channel.
The double-degenerate scenario of \citet{Shen+13} also has a rapid expulsion of mass
in a common envelope phase, 
though without a sufficient density and too distant to explain observed SNe~Ia;n.
It has also been suggested that a CO WD merging with the core of a post-AGB star
can create a variety of interaction timescales \citep[e.g.,][]{Soker13}.
In the core-collapse hypothesis, the ejections that create the CSM may be caused by
instabilities within the star or a binary companion; but whether this is relevant for
SNe~Ia;n is unclear, as \citet{Inserra+16} found that PTF11kx was the only SN~Ia-CSM for 
which they would not favor a core-collapse progenitor.
Characterizing the mass, location, and extent of the CSM in SNe~Ia;n can 
distinguish these scenarios from one another.

Determining the origin of the CSM in SNe~Ia;n and possibly thus the progenitors of 
91T-like SNe~Ia relies partially on increasing our sample of SNe~Ia;n with a systematic
observation strategy. 
From our analyses of PTF11kx, and now SN~2015cp, some guidelines for future efforts
to discover and characterize SNe~Ia;n emerge. 
The first suggestion is to ensure that nearby 91T-like SNe~Ia are monitored over a 
years-long baseline, to discover the onset of interaction:
our analysis shows that the two-year gap in observations of SN~2015cp severely limits 
our ability to use the radio nondetections {\it a priori}.
Long-term photometric monitoring is possible with the Zwicky Transient Facility  
(and, in the future, the Large Synoptic Survey Telescope) and discovery should be possible
with such surveys because \Hal\ holds the $R$-band magnitude constant months after 
SN peak \citep{WoodVasey04,Dilday+12}. 
Obtaining spectroscopic confirmation of the SN sub-type is also important and 
should be feasible if not rote for targets that are within 200 Mpc (which are the best-suited
targets for radio and X-ray follow-up).
After discovery of interaction, regular spectroscopy or imaging with a 
narrow \Hal\ filter to measure the \Hal\ line strength is strongly encouraged, 
as it is a proxy for the phase of interaction.
We have shown that radio observations should prioritize being {\it early} over being
{\it deep}: the AMI nondetection was nearly as constraining as that of the VLA. 
In the case of a radio detection, of course, the capability of the VLA to provide
a spectrum would be crucial and a time series useful. 
X-ray observations also constrain the CSM mass and are likely from thermal 
bremsstrahlung, thus independent of the synchrotron $\epsilon_B$ parameter.
If the \Hal\ emission is observed to be constant or increasing, 
a radio or X-ray nondetection may indicates a high optical depth in which case 
continued monitoring (or a different frequency observation) would be prudent.
Therefore, we suggest that continual monitoring of 91T-like SNe~Ia with a plan for rapid 
combined radio and optical follow-up observations is the path forward for growing our 
sample of SNe~Ia;n and understanding the progenitors of 91T-like SNe~Ia.

\section*{Acknowledgements}
A.H. acknowledges support by the I-Core Program of the Planning and Budgeting Committee
and the Israel Science Foundation. This research was supported by a Grant from the GIF, the German-Israeli Foundation for Scientific Research and Development. Support for A.V.F.'s supernova research group has been provided by the TABASGO Foundation, Gary and
Cynthia Bengier, the Christopher R. Redlich Fund, and
the Miller Institute for Basic Research in Science (U.C. Berkeley).
A.V.F.'s work was conducted in part at the Aspen Center for Physics,
which is supported by NSF grant PHY-1607611; he thanks the Center for
its hospitality during the supermassive black holes workshop in June
and July 2018.
K.M. acknowledges support from the UK STFC through an Ernest Rutherford Fellowship and from Horizon 2020 ERC Starting Grant (\#758638).
K.J.S. is supported by NASA through the Astrophysics Theory Program (NNX17AG28G).

\software{
reduce\_dc \citep{Perrott+13}, 
CASA \citep{CASA},
SciPy \citep{SciPy}, 
NumpPy \citep{NumPy}, 
Astropy \citep{Astropy},
Matplotlib \citep{Matplotlib}
}

\bibliographystyle{apj}
\bibliography{refs}

\begin{thebibliography}{47}
\expandafter\ifx\csname natexlab\endcsname\relax\def\natexlab#1{#1}\fi

\bibitem[{{Astropy Collaboration} {et~al.}(2013){Astropy Collaboration},
  {Robitaille}, {Tollerud}, {Greenfield}, {Droettboom}, {Bray}, {Aldcroft},
  {Davis}, {Ginsburg}, {Price-Whelan}, {Kerzendorf}, {Conley}, {Crighton},
  {Barbary}, {Muna}, {Ferguson}, {Grollier}, {Parikh}, {Nair}, {Unther},
  {Deil}, {Woillez}, {Conseil}, {Kramer}, {Turner}, {Singer}, {Fox}, {Weaver},
  {Zabalza}, {Edwards}, {Azalee Bostroem}, {Burke}, {Casey}, {Crawford},
  {Dencheva}, {Ely}, {Jenness}, {Labrie}, {Lim}, {Pierfederici}, {Pontzen},
  {Ptak}, {Refsdal}, {Servillat}, \& {Streicher}}]{Astropy}
{Astropy Collaboration}, {Robitaille}, T.~P., {Tollerud}, E.~J., {Greenfield},
  P., {Droettboom}, M., {Bray}, E., {Aldcroft}, T., {Davis}, M., {Ginsburg},
  A., {Price-Whelan}, A.~M., {Kerzendorf}, W.~E., {Conley}, A., {Crighton}, N.,
  {Barbary}, K., {Muna}, D., {Ferguson}, H., {Grollier}, F., {Parikh}, M.~M.,
  {Nair}, P.~H., {Unther}, H.~M., {Deil}, C., {Woillez}, J., {Conseil}, S.,
  {Kramer}, R., {Turner}, J.~E.~H., {Singer}, L., {Fox}, R., {Weaver}, B.~A.,
  {Zabalza}, V., {Edwards}, Z.~I., {Azalee Bostroem}, K., {Burke}, D.~J.,
  {Casey}, A.~R., {Crawford}, S.~M., {Dencheva}, N., {Ely}, J., {Jenness}, T.,
  {Labrie}, K., {Lim}, P.~L., {Pierfederici}, F., {Pontzen}, A., {Ptak}, A.,
  {Refsdal}, B., {Servillat}, M., \& {Streicher}, O. 2013, \aap, 558, A33

\bibitem[{{Borkowski} {et~al.}(2009){Borkowski}, {Blondin}, \&
  {Reynolds}}]{Borkowski+09}
{Borkowski}, K.~J., {Blondin}, J.~M., \& {Reynolds}, S.~P. 2009, \apjl, 699,
  L64

\bibitem[{{Branch} {et~al.}(1993){Branch}, {Fisher}, \& {Nugent}}]{Branch+93}
{Branch}, D., {Fisher}, A., \& {Nugent}, P. 1993, \aj, 106, 2383

\bibitem[{{Branch} {et~al.}(1995){Branch}, {Livio}, {Yungelson}, {Boffi}, \&
  {Baron}}]{Branch+95}
{Branch}, D., {Livio}, M., {Yungelson}, L.~R., {Boffi}, F.~R., \& {Baron}, E.
  1995, \pasp, 107, 1019

\bibitem[{{Bulla} {et~al.}(2018){Bulla}, {Goobar}, \& {Dhawan}}]{Bulla+18}
{Bulla}, M., {Goobar}, A., \& {Dhawan}, S. 2018, \mnras, 479, 3663

\bibitem[{{Chevalier}(1982)}]{Chevalier82a}
{Chevalier}, R.~A. 1982, \apj, 258, 790

\bibitem[{{Chomiuk} {et~al.}(2016){Chomiuk}, {Soderberg}, {Chevalier},
  {Bruzewski}, {Foley}, {Parrent}, {Strader}, {Badenes}, {Fransson}, {Kamble},
  {Margutti}, {Rupen}, \& {Simon}}]{Chomiuk+16}
{Chomiuk}, L., {Soderberg}, A.~M., {Chevalier}, R.~A., {Bruzewski}, S.,
  {Foley}, R.~J., {Parrent}, J., {Strader}, J., {Badenes}, C., {Fransson}, C.,
  {Kamble}, A., {Margutti}, R., {Rupen}, M.~P., \& {Simon}, J.~D. 2016, \apj,
  821, 119

\bibitem[{{Chugai}(2008)}]{Chugai08}
{Chugai}, N.~N. 2008, Astronomy Letters, 34, 389

\bibitem[{{Condon} {et~al.}(1998){Condon}, {Cotton}, {Greisen}, {Yin},
  {Perley}, {Taylor}, \& {Broderick}}]{Condon+98}
{Condon}, J.~J., {Cotton}, W.~D., {Greisen}, E.~W., {Yin}, Q.~F., {Perley},
  R.~A., {Taylor}, G.~B., \& {Broderick}, J.~J. 1998, \aj, 115, 1693

\bibitem[{{Dilday} {et~al.}(2012){Dilday}, {Howell}, {Cenko}, {Silverman},
  {Nugent}, {Sullivan}, {Ben-Ami}, {Bildsten}, {Bolte}, {Endl}, {Filippenko},
  {Gnat}, {Horesh}, {Hsiao}, {Kasliwal}, {Kirkman}, {Maguire}, {Marcy},
  {Moore}, {Pan}, {Parrent}, {Podsiadlowski}, {Quimby}, {Sternberg}, {Suzuki},
  {Tytler}, {Xu}, {Bloom}, {Gal-Yam}, {Hook}, {Kulkarni}, {Law}, {Ofek},
  {Polishook}, \& {Poznanski}}]{Dilday+12}
{Dilday}, B., {Howell}, D.~A., {Cenko}, S.~B., {Silverman}, J.~M., {Nugent},
  P.~E., {Sullivan}, M., {Ben-Ami}, S., {Bildsten}, L., {Bolte}, M., {Endl},
  M., {Filippenko}, A.~V., {Gnat}, O., {Horesh}, A., {Hsiao}, E., {Kasliwal},
  M.~M., {Kirkman}, D., {Maguire}, K., {Marcy}, G.~W., {Moore}, K., {Pan}, Y.,
  {Parrent}, J.~T., {Podsiadlowski}, P., {Quimby}, R.~M., {Sternberg}, A.,
  {Suzuki}, N., {Tytler}, D.~R., {Xu}, D., {Bloom}, J.~S., {Gal-Yam}, A.,
  {Hook}, I.~M., {Kulkarni}, S.~R., {Law}, N.~M., {Ofek}, E.~O., {Polishook},
  D., \& {Poznanski}, D. 2012, Science, 337, 942

\bibitem[{{Filippenko}(1997)}]{Filippenko97}
{Filippenko}, A.~V. 1997, \araa, 35, 309

\bibitem[{{Filippenko} {et~al.}(1992){Filippenko}, {Richmond}, {Matheson},
  {Shields}, {Burbidge}, {Cohen}, {Dickinson}, {Malkan}, {Nelson}, {Pietz},
  {Schlegel}, {Schmeer}, {Spinrad}, {Steidel}, {Tran}, \&
  {Wren}}]{Filippenko+92}
{Filippenko}, A.~V., {Richmond}, M.~W., {Matheson}, T., {Shields}, J.~C.,
  {Burbidge}, E.~M., {Cohen}, R.~D., {Dickinson}, M., {Malkan}, M.~A.,
  {Nelson}, B., {Pietz}, J., {Schlegel}, D., {Schmeer}, P., {Spinrad}, H.,
  {Steidel}, C.~C., {Tran}, H.~D., \& {Wren}, W. 1992, \apjl, 384, L15

\bibitem[{{Fink} {et~al.}(2010){Fink}, {R{\"o}pke}, {Hillebrandt},
  {Seitenzahl}, {Sim}, \& {Kromer}}]{Fink+10}
{Fink}, M., {R{\"o}pke}, F.~K., {Hillebrandt}, W., {Seitenzahl}, I.~R., {Sim},
  S.~A., \& {Kromer}, M. 2010, \aap, 514, A53

\bibitem[{{Fransson} {et~al.}(2015){Fransson}, {Larsson}, {Migotto}, {Pesce},
  {Challis}, {Chevalier}, {France}, {Kirshner}, {Leibundgut}, {Lundqvist},
  {McCray}, {Spyromilio}, {Taddia}, {Jerkstrand}, {Mattila}, {Smith},
  {Sollerman}, {Wheeler}, {Crotts}, {Garnavich}, {Heng}, {Lawrence}, {Panagia},
  {Pun}, {Sonneborn}, \& {Sugerman}}]{Fransson+15}
{Fransson}, C., {Larsson}, J., {Migotto}, K., {Pesce}, D., {Challis}, P.,
  {Chevalier}, R.~A., {France}, K., {Kirshner}, R.~P., {Leibundgut}, B.,
  {Lundqvist}, P., {McCray}, R., {Spyromilio}, J., {Taddia}, F., {Jerkstrand},
  A., {Mattila}, S., {Smith}, N., {Sollerman}, J., {Wheeler}, J.~C., {Crotts},
  A., {Garnavich}, P., {Heng}, K., {Lawrence}, S.~S., {Panagia}, N., {Pun},
  C.~S.~J., {Sonneborn}, G., \& {Sugerman}, B. 2015, \apjl, 806, L19

\bibitem[{{Garavini} {et~al.}(2004){Garavini}, {Folatelli}, {Goobar}, {Nobili},
  {Aldering}, {Amadon}, {Amanullah}, {Astier}, {Balland}, {Blanc}, {Burns},
  {Conley}, {Dahl{\'e}n}, {Deustua}, {Ellis}, {Fabbro}, {Fan}, {Frye}, {Gates},
  {Gibbons}, {Goldhaber}, {Goldman}, {Groom}, {Haissinski}, {Hardin}, {Hook},
  {Howell}, {Kasen}, {Kent}, {Kim}, {Knop}, {Lee}, {Lidman}, {Mendez},
  {Miller}, {Moniez}, {Mour{\~a}o}, {Newberg}, {Nugent}, {Pain}, {Perdereau},
  {Perlmutter}, {Prasad}, {Quimby}, {Raux}, {Regnault}, {Rich}, {Richards},
  {Ruiz-Lapuente}, {Sainton}, {Schaefer}, {Schahmaneche}, {Smith}, {Spadafora},
  {Stanishev}, {Walton}, {Wang}, {Wood-Vasey}, \& {Supernova Cosmology
  Project}}]{Garavini+04}
{Garavini}, G., {Folatelli}, G., {Goobar}, A., {Nobili}, S., {Aldering}, G.,
  {Amadon}, A., {Amanullah}, R., {Astier}, P., {Balland}, C., {Blanc}, G.,
  {Burns}, M.~S., {Conley}, A., {Dahl{\'e}n}, T., {Deustua}, S.~E., {Ellis},
  R., {Fabbro}, S., {Fan}, X., {Frye}, B., {Gates}, E.~L., {Gibbons}, R.,
  {Goldhaber}, G., {Goldman}, B., {Groom}, D.~E., {Haissinski}, J., {Hardin},
  D., {Hook}, I.~M., {Howell}, D.~A., {Kasen}, D., {Kent}, S., {Kim}, A.~G.,
  {Knop}, R.~A., {Lee}, B.~C., {Lidman}, C., {Mendez}, J., {Miller}, G.~J.,
  {Moniez}, M., {Mour{\~a}o}, A., {Newberg}, H., {Nugent}, P.~E., {Pain}, R.,
  {Perdereau}, O., {Perlmutter}, S., {Prasad}, V., {Quimby}, R., {Raux}, J.,
  {Regnault}, N., {Rich}, J., {Richards}, G.~T., {Ruiz-Lapuente}, P.,
  {Sainton}, G., {Schaefer}, B.~E., {Schahmaneche}, K., {Smith}, E.,
  {Spadafora}, A.~L., {Stanishev}, V., {Walton}, N.~A., {Wang}, L.,
  {Wood-Vasey}, W.~M., \& {Supernova Cosmology Project}. 2004, \aj, 128, 387

\bibitem[{{Graham} {et~al.}(2017){Graham}, {Harris}, {Fox}, {Nugent}, {Kasen},
  {Silverman}, \& {Filippenko}}]{GH+17}
{Graham}, M.~L., {Harris}, C.~E., {Fox}, O.~D., {Nugent}, P.~E., {Kasen}, D.,
  {Silverman}, J.~M., \& {Filippenko}, A.~V. 2017, \apj, 843, 102

\bibitem[{{Graham} {et~al.}(2018){Graham}, {Harris}, {Nugent}, {Maguire},
  {Sullivan}, {Smith}, {Valenti}, {Goobar}, {Fox}, {Shen}, {Kelly}, McCully,
  {Brink}, \& {Filippenko}}]{Graham+18}
{Graham}, M.~L., {Harris}, C.~E., {Nugent}, P.~E., {Maguire}, K., {Sullivan},
  M., {Smith}, M., {Valenti}, S., {Goobar}, A., {Fox}, O.~D., {Shen}, K.~J.,
  {Kelly}, P.~L., McCully, C., {Brink}, T.~G., \& {Filippenko}, A.~V. 2018,
  \apj, in review

\bibitem[{{Harris} {et~al.}(2016){Harris}, {Nugent}, \& {Kasen}}]{HNK16}
{Harris}, C.~E., {Nugent}, P.~E., \& {Kasen}, D.~N. 2016, \apj, 823, 100

\bibitem[{{Hickish} {et~al.}(2018){Hickish}, {Razavi-Ghods}, {Perrott},
  {Titterington}, {Carey}, {Scott}, {Grainge}, {Scaife}, {Alexander},
  {Saunders}, {Crofts}, {Javid}, {Rumsey}, {Jin}, {Ely}, {Shaw}, {Northrop},
  {Pooley}, {D'Alessandro}, {Doherty}, \& {Willatt}}]{Hickish+18}
{Hickish}, J., {Razavi-Ghods}, N., {Perrott}, Y.~C., {Titterington}, D.~J.,
  {Carey}, S.~H., {Scott}, P.~F., {Grainge}, K.~J.~B., {Scaife}, A.~M.~M.,
  {Alexander}, P., {Saunders}, R.~D.~E., {Crofts}, M., {Javid}, K., {Rumsey},
  C., {Jin}, T.~Z., {Ely}, J.~A., {Shaw}, C., {Northrop}, I.~G., {Pooley}, G.,
  {D'Alessandro}, R., {Doherty}, P., \& {Willatt}, G.~P. 2018, \mnras, 475,
  5677

\bibitem[{{Hunter}(2007)}]{Matplotlib}
{Hunter}, J.~D. 2007, Computing in Science \& Engineering, 9, 90

\bibitem[{{Inserra} {et~al.}(2016){Inserra}, {Fraser}, {Smartt}, {Benetti},
  {Chen}, {Childress}, {Gal-Yam}, {Howell}, {Kangas}, {Pignata}, {Polshaw},
  {Sullivan}, {Smith}, {Valenti}, {Young}, {Parker}, {Seccull}, \&
  {McCrum}}]{Inserra+16}
{Inserra}, C., {Fraser}, M., {Smartt}, S.~J., {Benetti}, S., {Chen}, T.-W.,
  {Childress}, M., {Gal-Yam}, A., {Howell}, D.~A., {Kangas}, T., {Pignata}, G.,
  {Polshaw}, J., {Sullivan}, M., {Smith}, K.~W., {Valenti}, S., {Young}, D.~R.,
  {Parker}, S., {Seccull}, T., \& {McCrum}, M. 2016, \mnras, 459, 2721

\bibitem[{{Jones} {et~al.}(2001){Jones}, {Oliphant}, {Peterson},
  {et~al.}}]{SciPy}
{Jones}, E., {Oliphant}, T., {Peterson}, P., {et~al.} 2001, {SciPy}: Open
  source scientific tools for {Python}, [Online; accessed <today>]

\bibitem[{{Kalberla} {et~al.}(2005){Kalberla}, {Burton}, {Hartmann}, {Arnal},
  {Bajaja}, {Morras}, \& {P{\"o}ppel}}]{Kalberla+05}
{Kalberla}, P.~M.~W., {Burton}, W.~B., {Hartmann}, D., {Arnal}, E.~M.,
  {Bajaja}, E., {Morras}, R., \& {P{\"o}ppel}, W.~G.~L. 2005, \aap, 440, 775

\bibitem[{{Larsson} {et~al.}(2011){Larsson}, {Fransson}, {{\"O}stlin},
  {Gr{\"o}ningsson}, {Jerkstrand}, {Kozma}, {Sollerman}, {Challis}, {Kirshner},
  {Chevalier}, {Heng}, {McCray}, {Suntzeff}, {Bouchet}, {Crotts}, {Danziger},
  {Dwek}, {France}, {Garnavich}, {Lawrence}, {Leibundgut}, {Lundqvist},
  {Panagia}, {Pun}, {Smith}, {Sonneborn}, {Wang}, \& {Wheeler}}]{Larsson+11}
{Larsson}, J., {Fransson}, C., {{\"O}stlin}, G., {Gr{\"o}ningsson}, P.,
  {Jerkstrand}, A., {Kozma}, C., {Sollerman}, J., {Challis}, P., {Kirshner},
  R.~P., {Chevalier}, R.~A., {Heng}, K., {McCray}, R., {Suntzeff}, N.~B.,
  {Bouchet}, P., {Crotts}, A., {Danziger}, J., {Dwek}, E., {France}, K.,
  {Garnavich}, P.~M., {Lawrence}, S.~S., {Leibundgut}, B., {Lundqvist}, P.,
  {Panagia}, N., {Pun}, C.~S.~J., {Smith}, N., {Sonneborn}, G., {Wang}, L., \&
  {Wheeler}, J.~C. 2011, \nat, 474, 484

\bibitem[{{Leloudas} {et~al.}(2015){Leloudas}, {Hsiao}, {Johansson}, {Maeda},
  {Moriya}, {Nordin}, {Petrushevska}, {Silverman}, {Sollerman}, {Stritzinger},
  {Taddia}, \& {Xu}}]{Leloudas+15}
{Leloudas}, G., {Hsiao}, E.~Y., {Johansson}, J., {Maeda}, K., {Moriya}, T.~J.,
  {Nordin}, J., {Petrushevska}, T., {Silverman}, J.~M., {Sollerman}, J.,
  {Stritzinger}, M.~D., {Taddia}, F., \& {Xu}, D. 2015, \aap, 574, A61

\bibitem[{{Maguire} {et~al.}(2013){Maguire}, {Sullivan}, {Patat}, {Gal-Yam},
  {Hook}, {Dhawan}, {Howell}, {Mazzali}, {Nugent}, {Pan}, {Podsiadlowski},
  {Simon}, {Sternberg}, {Valenti}, {Baltay}, {Bersier}, {Blagorodnova}, {Chen},
  {Ellman}, {Feindt}, {F{\"o}rster}, {Fraser}, {Gonz{\'a}lez-Gait{\'a}n},
  {Graham}, {Guti{\'e}rrez}, {Hachinger}, {Hadjiyska}, {Inserra}, {Knapic},
  {Laher}, {Leloudas}, {Margheim}, {McKinnon}, {Molinaro}, {Morrell}, {Ofek},
  {Rabinowitz}, {Rest}, {Sand}, {Smareglia}, {Smartt}, {Taddia}, {Walker},
  {Walton}, \& {Young}}]{Maguire+13}
{Maguire}, K., {Sullivan}, M., {Patat}, F., {Gal-Yam}, A., {Hook}, I.~M.,
  {Dhawan}, S., {Howell}, D.~A., {Mazzali}, P., {Nugent}, P.~E., {Pan}, Y.-C.,
  {Podsiadlowski}, P., {Simon}, J.~D., {Sternberg}, A., {Valenti}, S.,
  {Baltay}, C., {Bersier}, D., {Blagorodnova}, N., {Chen}, T.-W., {Ellman}, N.,
  {Feindt}, U., {F{\"o}rster}, F., {Fraser}, M., {Gonz{\'a}lez-Gait{\'a}n}, S.,
  {Graham}, M.~L., {Guti{\'e}rrez}, C., {Hachinger}, S., {Hadjiyska}, E.,
  {Inserra}, C., {Knapic}, C., {Laher}, R.~R., {Leloudas}, G., {Margheim}, S.,
  {McKinnon}, R., {Molinaro}, M., {Morrell}, N., {Ofek}, E.~O., {Rabinowitz},
  D., {Rest}, A., {Sand}, D., {Smareglia}, R., {Smartt}, S.~J., {Taddia}, F.,
  {Walker}, E.~S., {Walton}, N.~A., \& {Young}, D.~R. 2013, \mnras, 436, 222

\bibitem[{{Mauerhan} {et~al.}(2018){Mauerhan}, {Filippenko}, {Zheng}, {Brink},
  {Graham}, {Shivvers}, \& {Clubb}}]{Mauerhan+18}
{Mauerhan}, J.~C., {Filippenko}, A.~V., {Zheng}, W., {Brink}, T.~G., {Graham},
  M.~L., {Shivvers}, I., \& {Clubb}, K.~I. 2018, \mnras, 478, 5050

\bibitem[{{McMullin} {et~al.}(2007){McMullin}, {Waters}, {Schiebel}, {Young},
  \& {Golap}}]{CASA}
{McMullin}, J.~P., {Waters}, B., {Schiebel}, D., {Young}, W., \& {Golap}, K.
  2007, in Astronomical Society of the Pacific Conference Series, Vol. 376,
  Astronomical Data Analysis Software and Systems XVI, ed. R.~A. {Shaw},
  F.~{Hill}, \& D.~J. {Bell}, 127

\bibitem[{{Milisavljevic} {et~al.}(2015){Milisavljevic}, {Margutti}, {Kamble},
  {Patnaude}, {Raymond}, {Eldridge}, {Fong}, {Bietenholz}, {Challis},
  {Chornock}, {Drout}, {Fransson}, {Fesen}, {Grindlay}, {Kirshner}, {Lunnan},
  {Mackey}, {Miller}, {Parrent}, {Sanders}, {Soderberg}, \&
  {Zauderer}}]{Milisav+15}
{Milisavljevic}, D., {Margutti}, R., {Kamble}, A., {Patnaude}, D.~J.,
  {Raymond}, J.~C., {Eldridge}, J.~J., {Fong}, W., {Bietenholz}, M., {Challis},
  P., {Chornock}, R., {Drout}, M.~R., {Fransson}, C., {Fesen}, R.~A.,
  {Grindlay}, J.~E., {Kirshner}, R.~P., {Lunnan}, R., {Mackey}, J., {Miller},
  G.~F., {Parrent}, J.~T., {Sanders}, N.~E., {Soderberg}, A.~M., \& {Zauderer},
  B.~A. 2015, \apj, 815, 120

\bibitem[{{Moore} \& {Bildsten}(2012)}]{MooreBildsten12}
{Moore}, K. \& {Bildsten}, L. 2012, \apj, 761, 182

\bibitem[{{O'Brien} {et~al.}(1992){O'Brien}, {Bode}, \& {Kahn}}]{OBrien+92}
{O'Brien}, T.~J., {Bode}, M.~F., \& {Kahn}, F.~D. 1992, \mnras, 255, 683

\bibitem[{{Oliphant}(2006)}]{NumPy}
{Oliphant}, T. 2006, {A guide to NumPy} (USA, Trelgol Publishing)

\bibitem[{{Patat} {et~al.}(2007){Patat}, {Chandra}, {Chevalier}, {Justham},
  {Podsiadlowski}, {Wolf}, {Gal-Yam}, {Pasquini}, {Crawford}, {Mazzali},
  {Pauldrach}, {Nomoto}, {Benetti}, {Cappellaro}, {Elias-Rosa}, {Hillebrandt},
  {Leonard}, {Pastorello}, {Renzini}, {Sabbadin}, {Simon}, \&
  {Turatto}}]{Patat+07}
{Patat}, F., {Chandra}, P., {Chevalier}, R., {Justham}, S., {Podsiadlowski},
  P., {Wolf}, C., {Gal-Yam}, A., {Pasquini}, L., {Crawford}, I.~A., {Mazzali},
  P.~A., {Pauldrach}, A.~W.~A., {Nomoto}, K., {Benetti}, S., {Cappellaro}, E.,
  {Elias-Rosa}, N., {Hillebrandt}, W., {Leonard}, D.~C., {Pastorello}, A.,
  {Renzini}, A., {Sabbadin}, F., {Simon}, J.~D., \& {Turatto}, M. 2007,
  Science, 317, 924

\bibitem[{{Perrott} {et~al.}(2013){Perrott}, {Scaife}, {Green}, {Davies},
  {Franzen}, {Grainge}, {Hobson}, {Hurley-Walker}, {Lasenby}, {Olamaie},
  {Pooley}, {Rodr{\'{\i}}guez-Gonz{\'a}lvez}, {Rumsey}, {Saunders}, {Schammel},
  {Scott}, {Shimwell}, {Titterington}, {Waldram}, \& {AMI
  Consortium}}]{Perrott+13}
{Perrott}, Y.~C., {Scaife}, A.~M.~M., {Green}, D.~A., {Davies}, M.~L.,
  {Franzen}, T.~M.~O., {Grainge}, K.~J.~B., {Hobson}, M.~P., {Hurley-Walker},
  N., {Lasenby}, A.~N., {Olamaie}, M., {Pooley}, G.~G.,
  {Rodr{\'{\i}}guez-Gonz{\'a}lvez}, C., {Rumsey}, C., {Saunders}, R.~D.~E.,
  {Schammel}, M.~P., {Scott}, P.~F., {Shimwell}, T.~W., {Titterington}, D.~J.,
  {Waldram}, E.~M., \& {AMI Consortium}. 2013, \mnras, 429, 3330

\bibitem[{{Raskin} \& {Kasen}(2013)}]{RaskinKasen13}
{Raskin}, C. \& {Kasen}, D. 2013, \apj, 772, 1

\bibitem[{{Shen} {et~al.}(2013){Shen}, {Guillochon}, \& {Foley}}]{Shen+13}
{Shen}, K.~J., {Guillochon}, J., \& {Foley}, R.~J. 2013, \apjl, 770, L35

\bibitem[{{Shen} \& {Moore}(2014)}]{ShenMoore14}
{Shen}, K.~J. \& {Moore}, K. 2014, \apj, 797, 46

\bibitem[{{Silverman} {et~al.}(2013{\natexlab{a}}){Silverman}, {Nugent},
  {Gal-Yam}, {Sullivan}, {Howell}, {Filippenko}, {Arcavi}, {Ben-Ami}, {Bloom},
  {Cenko}, {Cao}, {Chornock}, {Clubb}, {Coil}, {Foley}, {Graham}, {Griffith},
  {Horesh}, {Kasliwal}, {Kulkarni}, {Leonard}, {Li}, {Matheson}, {Miller},
  {Modjaz}, {Ofek}, {Pan}, {Perley}, {Poznanski}, {Quimby}, {Steele},
  {Sternberg}, {Xu}, \& {Yaron}}]{Silverman+13}
{Silverman}, J.~M., {Nugent}, P.~E., {Gal-Yam}, A., {Sullivan}, M., {Howell},
  D.~A., {Filippenko}, A.~V., {Arcavi}, I., {Ben-Ami}, S., {Bloom}, J.~S.,
  {Cenko}, S.~B., {Cao}, Y., {Chornock}, R., {Clubb}, K.~I., {Coil}, A.~L.,
  {Foley}, R.~J., {Graham}, M.~L., {Griffith}, C.~V., {Horesh}, A., {Kasliwal},
  M.~M., {Kulkarni}, S.~R., {Leonard}, D.~C., {Li}, W., {Matheson}, T.,
  {Miller}, A.~A., {Modjaz}, M., {Ofek}, E.~O., {Pan}, Y.-C., {Perley}, D.~A.,
  {Poznanski}, D., {Quimby}, R.~M., {Steele}, T.~N., {Sternberg}, A., {Xu}, D.,
  \& {Yaron}, O. 2013{\natexlab{a}}, \apjs, 207, 3

\bibitem[{{Silverman} {et~al.}(2013{\natexlab{b}}){Silverman}, {Nugent},
  {Gal-Yam}, {Sullivan}, {Howell}, {Filippenko}, {Pan}, {Cenko}, \&
  {Hook}}]{Silverman+13b}
{Silverman}, J.~M., {Nugent}, P.~E., {Gal-Yam}, A., {Sullivan}, M., {Howell},
  D.~A., {Filippenko}, A.~V., {Pan}, Y.-C., {Cenko}, S.~B., \& {Hook}, I.~M.
  2013{\natexlab{b}}, \apj, 772, 125

\bibitem[{{Simon} {et~al.}(2009){Simon}, {Gal-Yam}, {Gnat}, {Quimby},
  {Ganeshalingam}, {Silverman}, {Blondin}, {Li}, {Filippenko}, {Wheeler},
  {Kirshner}, {Patat}, {Nugent}, {Foley}, {Vogt}, {Butler}, {Peek},
  {Rosolowsky}, {Herczeg}, {Sauer}, \& {Mazzali}}]{Simon+09}
{Simon}, J.~D., {Gal-Yam}, A., {Gnat}, O., {Quimby}, R.~M., {Ganeshalingam},
  M., {Silverman}, J.~M., {Blondin}, S., {Li}, W., {Filippenko}, A.~V.,
  {Wheeler}, J.~C., {Kirshner}, R.~P., {Patat}, F., {Nugent}, P., {Foley},
  R.~J., {Vogt}, S.~S., {Butler}, R.~P., {Peek}, K.~M.~G., {Rosolowsky}, E.,
  {Herczeg}, G.~J., {Sauer}, D.~N., \& {Mazzali}, P.~A. 2009, \apj, 702, 1157

\bibitem[{{Soker}(2013)}]{Soker13}
{Soker}, N. 2013, in IAU Symposium, Vol. 281, Binary Paths to Type Ia
  Supernovae Explosions, ed. R.~{Di Stefano}, M.~{Orio}, \& M.~{Moe}, 72--75

\bibitem[{{Sternberg} {et~al.}(2011){Sternberg}, {Gal-Yam}, {Simon}, {Leonard},
  {Quimby}, {Phillips}, {Morrell}, {Thompson}, {Ivans}, {Marshall},
  {Filippenko}, {Marcy}, {Bloom}, {Patat}, {Foley}, {Yong}, {Penprase},
  {Beeler}, {Allende Prieto}, \& {Stringfellow}}]{Sternberg+11}
{Sternberg}, A., {Gal-Yam}, A., {Simon}, J.~D., {Leonard}, D.~C., {Quimby},
  R.~M., {Phillips}, M.~M., {Morrell}, N., {Thompson}, I.~B., {Ivans}, I.,
  {Marshall}, J.~L., {Filippenko}, A.~V., {Marcy}, G.~W., {Bloom}, J.~S.,
  {Patat}, F., {Foley}, R.~J., {Yong}, D., {Penprase}, B.~E., {Beeler}, D.~J.,
  {Allende Prieto}, C., \& {Stringfellow}, G.~S. 2011, Science, 333, 856

\bibitem[{{Sternberg} {et~al.}(2014){Sternberg}, {Gal-Yam}, {Simon}, {Patat},
  {Hillebrandt}, {Phillips}, {Foley}, {Thompson}, {Morrell}, {Chomiuk},
  {Soderberg}, {Yong}, {Kraus}, {Herczeg}, {Hsiao}, {Raskutti}, {Cohen},
  {Mazzali}, \& {Nomoto}}]{Sternberg+14}
{Sternberg}, A., {Gal-Yam}, A., {Simon}, J.~D., {Patat}, F., {Hillebrandt}, W.,
  {Phillips}, M.~M., {Foley}, R.~J., {Thompson}, I., {Morrell}, N., {Chomiuk},
  L., {Soderberg}, A.~M., {Yong}, D., {Kraus}, A.~L., {Herczeg}, G.~J.,
  {Hsiao}, E.~Y., {Raskutti}, S., {Cohen}, J.~G., {Mazzali}, P.~A., \&
  {Nomoto}, K. 2014, \mnras, 443, 1849

\bibitem[{{van Hoof} {et~al.}(2014){van Hoof}, {Williams}, {Volk}, {Chatzikos},
  {Ferland}, {Lykins}, {Porter}, \& {Wang}}]{vanHoof+14}
{van Hoof}, P.~A.~M., {Williams}, R.~J.~R., {Volk}, K., {Chatzikos}, M.,
  {Ferland}, G.~J., {Lykins}, M., {Porter}, R.~L., \& {Wang}, Y. 2014, \mnras,
  444, 420

\bibitem[{{Wang} {et~al.}(2008){Wang}, {Li}, {Filippenko}, {Krisciunas},
  {Suntzeff}, {Li}, {Zhang}, {Deng}, {Foley}, {Ganeshalingam}, {Li}, {Lou},
  {Qiu}, {Shang}, {Silverman}, {Zhang}, \& {Zhang}}]{Wang+08}
{Wang}, X., {Li}, W., {Filippenko}, A.~V., {Krisciunas}, K., {Suntzeff}, N.~B.,
  {Li}, J., {Zhang}, T., {Deng}, J., {Foley}, R.~J., {Ganeshalingam}, M., {Li},
  T., {Lou}, Y., {Qiu}, Y., {Shang}, R., {Silverman}, J.~M., {Zhang}, S., \&
  {Zhang}, Y. 2008, \apj, 675, 626

\bibitem[{{Wood-Vasey} {et~al.}(2004){Wood-Vasey}, {Wang}, \&
  {Aldering}}]{WoodVasey04}
{Wood-Vasey}, W.~M., {Wang}, L., \& {Aldering}, G. 2004, \apj, 616, 339

\bibitem[{{Zwart} {et~al.}(2008){Zwart}, {Barker}, {Biddulph}, {Bly}, {Boysen},
  {Brown}, {Clementson}, {Crofts}, {Culverhouse}, {Czeres}, {Dace}, {Davies},
  {D'Alessandro}, {Doherty}, {Duggan}, {Ely}, {Felvus}, {Feroz}, {Flynn},
  {Franzen}, {Geisb{\"u}sch}, {G{\'e}nova-Santos}, {Grainge}, {Grainger},
  {Hammett}, {Hills}, {Hobson}, {Holler}, {Hurley-Walker}, {Jilley}, {Jones},
  {Kaneko}, {Kneissl}, {Lancaster}, {Lasenby}, {Marshall}, {Newton}, {Norris},
  {Northrop}, {Odell}, {Petencin}, {Pober}, {Pooley}, {Pospieszalski}, {Quy},
  {Rodr{\'{\i}}guez-Gonz{\'a}lvez}, {Saunders}, {Scaife}, {Schofield}, {Scott},
  {Shaw}, {Shimwell}, {Smith}, {Taylor}, {Titterington}, {Veli{\'c}},
  {Waldram}, {West}, {Wood}, {Yassin}, \& {AMI Consortium}}]{Zwart+08}
{Zwart}, J.~T.~L., {Barker}, R.~W., {Biddulph}, P., {Bly}, D., {Boysen}, R.~C.,
  {Brown}, A.~R., {Clementson}, C., {Crofts}, M., {Culverhouse}, T.~L.,
  {Czeres}, J., {Dace}, R.~J., {Davies}, M.~L., {D'Alessandro}, R., {Doherty},
  P., {Duggan}, K., {Ely}, J.~A., {Felvus}, M., {Feroz}, F., {Flynn}, W.,
  {Franzen}, T.~M.~O., {Geisb{\"u}sch}, J., {G{\'e}nova-Santos}, R., {Grainge},
  K.~J.~B., {Grainger}, W.~F., {Hammett}, D., {Hills}, R.~E., {Hobson}, M.~P.,
  {Holler}, C.~M., {Hurley-Walker}, N., {Jilley}, R., {Jones}, M.~E., {Kaneko},
  T., {Kneissl}, R., {Lancaster}, K., {Lasenby}, A.~N., {Marshall}, P.~J.,
  {Newton}, F., {Norris}, O., {Northrop}, I., {Odell}, D.~M., {Petencin}, G.,
  {Pober}, J.~C., {Pooley}, G.~G., {Pospieszalski}, M.~W., {Quy}, V.,
  {Rodr{\'{\i}}guez-Gonz{\'a}lvez}, C., {Saunders}, R.~D.~E., {Scaife},
  A.~M.~M., {Schofield}, J., {Scott}, P.~F., {Shaw}, C., {Shimwell}, T.~W.,
  {Smith}, H., {Taylor}, A.~C., {Titterington}, D.~J., {Veli{\'c}}, M.,
  {Waldram}, E.~M., {West}, S., {Wood}, B.~A., {Yassin}, G., \& {AMI
  Consortium}. 2008, \mnras, 391, 1545

\end{thebibliography}

\end{document}